\PassOptionsToPackage{
  paper=a4paper,
  left=0.8in,
  right=0.8in,
  top=0.8in,
  bottom=0.8in,
  twoside=false
}{geometry}

\documentclass[pdflatex,sn-mathphys-num]{sn-jnl}

\usepackage{setspace}
\usepackage{graphicx}%
\usepackage{multirow}%
\usepackage{amsmath,amssymb,amsfonts}%
\usepackage{amsthm}%
\usepackage{mathrsfs}%
\usepackage[title]{appendix}%
\usepackage{xcolor}%
\usepackage{textcomp}%
\usepackage{manyfoot}%
\usepackage{booktabs}%
\usepackage{algorithm}%
\usepackage{algorithmicx}%
\usepackage{algpseudocode}%
\usepackage{listings}%
\usepackage{tikz}
\usepackage{caption}
\DeclareCaptionFont{mid}{\fontsize{8}{11.5}\selectfont}
\theoremstyle{thmstyleone}%
\theoremstyle{thmstyletwo}%

\theoremstyle{thmstylethree}%

\begin{document}
\onehalfspacing
\title[Article Title]{Decoherence-Induced Multiphoton Interference}


\author[1,2]{\fnm{Yifan} \sur{Du}}\email{yifanduzd@gmail.com}

\author[1,2]{\fnm{Jiuyi} \sur{Zhang}}

\author[1,2]{\fnm{Daniel} \sur{López Martínez}}

\author[1,2]{\fnm{Santosh} \sur{Kumar}}

\author[1,2]{\fnm{Misagh} \sur{Izadi}}

\author*[1,2]{\fnm{Yuping} \sur{Huang}}\email{yhuang5@stevens.edu}

\affil*[1]{\orgdiv{Department of Physics}, \orgname{Stevens Institute of Technology}, \orgaddress{\street{1 Castle Point Terrace}, \city{Hoboken}, \postcode{07030}, \state{NJ}, \country{USA}}}

\affil[2]{\orgdiv{Center for Quantum Science and Engineering}, \orgname{Stevens Institute of Technology}, \orgaddress{\street{1 Castle Point Terrace}, \city{Hoboken}, \postcode{07030}, \state{NJ}, \country{USA}}}



\abstract{Decoherence is usually deemed detrimental to quantum information processing. Its control and minimization require significant costs and operating overheads, constituting a major hurdle to the practical deployment of quantum technologies. Yet, quantum mechanics provides for counterintuitive, sometimes surprisingly useful, phenomena and effects associated with decoherence, leading to unusual practical utilities. Here we demonstrate such an example of fundamental interest and practical potential, where quantum interference is created among multiple photons through their
dissipative coupling to a shared reservoir. On a thin-film lithium niobate chip, we dissipatively link two spontaneous parametric down-converters through a common, highly lossy channel to create phase-sensitive multiphoton states. Our results show that faithful correlations can be established among two, three, and four photons, and tuned by shifting the relative phase between the driving pumps for the converters. This experiment highlights an under-explored territory in quantum science and technology, where loss and decoherence serve as resources, rather than adversaries, for quantum information processing. 

}



\maketitle

\section{Introduction}\label{sec1}
In quantum information science, loss and decoherence are usually to be minimized or avoided, since they can cause quantum states to collapse, quantum features to degrade, and quantum advantages to vanish \cite{Q_compute_1,QP6}. However, more and more studies in open quantum systems have shown that dissipation can in fact be engineered to serve as an engine that drives a system toward a desirable quantum state, enabling, for example, the robust preparation of entangled states regardless of the initial condition and perturbations from the environment \cite{decoherence,DFS,loss_1,loss_2,loss_3,loss_4,loss_5,loss_6,loss_11}. Also, dissipative coupling through a common reservoir can link otherwise distinct quantum pathways and generate interference mechanisms in manners not possible with closed, coherent systems described by Hermitian Hamiltonians \cite{loss_7,loss_8,loss_9,loss_10}. 

In many aspects, such open-system paradigms are intrinsically connected to the broader framework of non-Hermitian physics. In recent studies, the fundamental axiom of the Hamiltonian being Hermitian as a mathematical constraint has been weakened by parity–time  (PT) symmetry as a physically transparent condition in quantum mechanics. Although non-Hermitian, a PT-symmetric Hamiltonian is sufficient to guarantee real eigenvalues and probability conservation \cite{bender,bender2}. This relaxation of constraints has inspired rich theoretical developments \cite{PT_theory1,PT_theory2,PT_theory3}, and some experimental studies \cite{PT_exp1,PT_exp2,PT_exp3}. Next to PT, anti-PT systems are characterized by Hamiltonians that anti-commute with the PT operator and exhibit purely imaginary eigenvalues in the unbroken phase \cite{Anti_PT1,Anti_PT2}. Unlike typical PT-symmetric systems where each component couples to an independent bath for gain and loss, an anti-PT system can be realized via dissipative coupling to a common reservoir \cite{Anti_PT3}. As such, anti-PT symmetry can serve as a promising candidate for studies in the quantum regime, as a robust means to generate and process quantum states without as much concern of noise injection. Furthermore, as an effective symmetry description of reservoir-engineered open-system dynamics, the anti-PT symmetry framework gives rise to rich spectral structures and phase transition phenomena \cite{QP4}. 

Despite significant interest and extensive theoretical studies, most experimental studies thus far have used classical systems to realize effective Hamiltonians with PT or anti-PT symmetry. Hence, they can only probe the physics on the level of first quantization, leaving many intrinsic quantum properties, including those of non-locality and non-realism, largely unexplored \cite{PT_1stquanti_1,PT_1stquanti_2,PT_1stquanti_3, Anti_PT1,Anti_PT2,Anti_PT3,Anti_PT4,Dirac}. Only recently, the effects of PT symmetry in second quantization have been experimentally studied through two-photon quantum interference \cite{2nd_quanti_0,2nd_quanti_1,2nd_quanti_2,2nd_quanti_3}, demonstrating counterintuitive effects like a shift of the Hong–Ou–Mandel (HOM) dip toward shorter interaction lengths and order-invariant correlations \cite{2nd_quanti_0,four_photon6}. Interestingly, quantum jumps are found to play a significant role in these regimes, a key quantum effect not captured in those classical systems \cite{QP1,QP2,QP3,QP4}.


Here, we push the studies of non-Hermitian quantum physics on the second quantization level beyond the biphoton interference and demonstrate, for the first time, multiphoton interference in the anti-PT regime. Among multiple platforms of choice \cite{QPT1,QPT2,QPT3},  photonic systems stand out as they operate at room temperature, provide device scalability, and allow repeatable engineering \cite{photonics_1,photonics_2}. Also, they can host nontrivial multiphoton quantum states that carry valuable, often much richer physics than that of two particles \cite{whyfour1,whyfour2,whyfour3,whyfour4}. Using photonics, we study (i) multiphoton correlations in the anti-PT (or PT) symmetry regime where the Hermiticity is relaxed; (ii) the effect of continuous environmental measurement—described by quantum jumps—on multiphoton quantum interference in open quantum systems \cite{QP5}; (iii) frustrated multiphoton quantum interference, in which nonlinearity is hybridized with interference via the dissipative superposition of indistinguishable photon-pair generation pathways  \cite{four_photon1,four_photon4,four_photon5,four_photon7}. 
On the application side, our demonstration points to a robust method to generate nontrivial multiphoton states for quantum computing, sensing, and other applications \cite{four_photon2}. 



Our system consists of two dissipatively coupled waveguides for spontaneous parametric down-conversion (SPDC) on a thin-film lithium niobate (TFLN) nanophotonic chip. Unlike those via coherent coupling, a system of such exhibits intrinsic robustness arising from a protected dark state to enable robust quantum-state generation. This also provides a paradigm to study coherent, non-classical dynamics under continuous  incoherent joint measurement on the second-quantization level, and the potential for exotic quantum information processing. Finally, the chip outputs nontrivial multiphoton correlations, with quantum interference on the second-, third-, and fourth-order levels. Our results illuminate the irreversible nature of quantum measurement in higher-order interference, and reveal how engineered loss can generate and control higher-order quantum interference, offering new insights into the quantum foundations of PT-symmetric systems and their applications. 






\section{Theoretical Model and Implementation}\label{sec2}

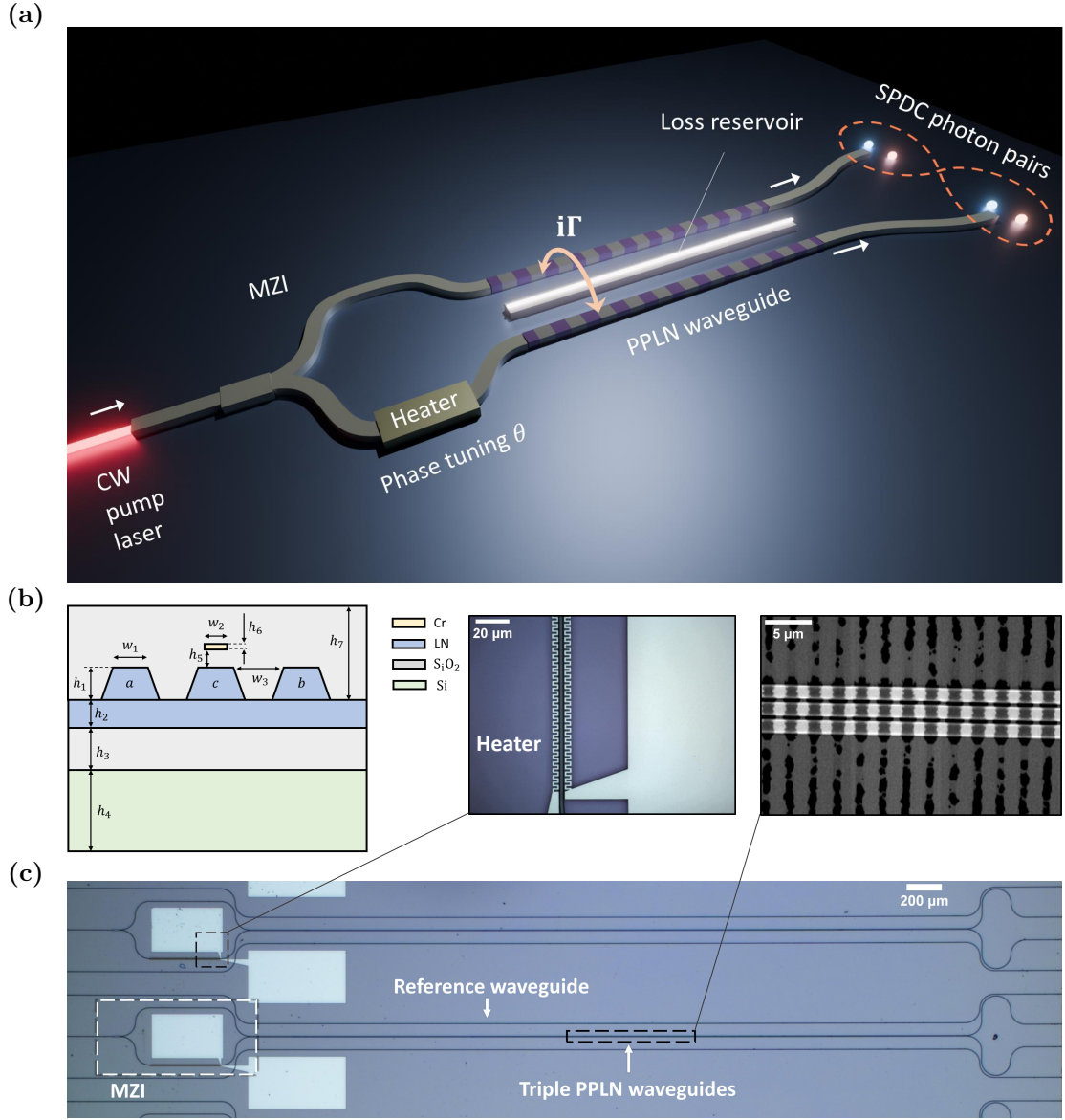
\begin{figure}[h!]
\centering
\begin{tikzpicture}
\node[anchor=south west, inner sep=0] (img) at (0,0)
    {\includegraphics[width=0.96\linewidth]{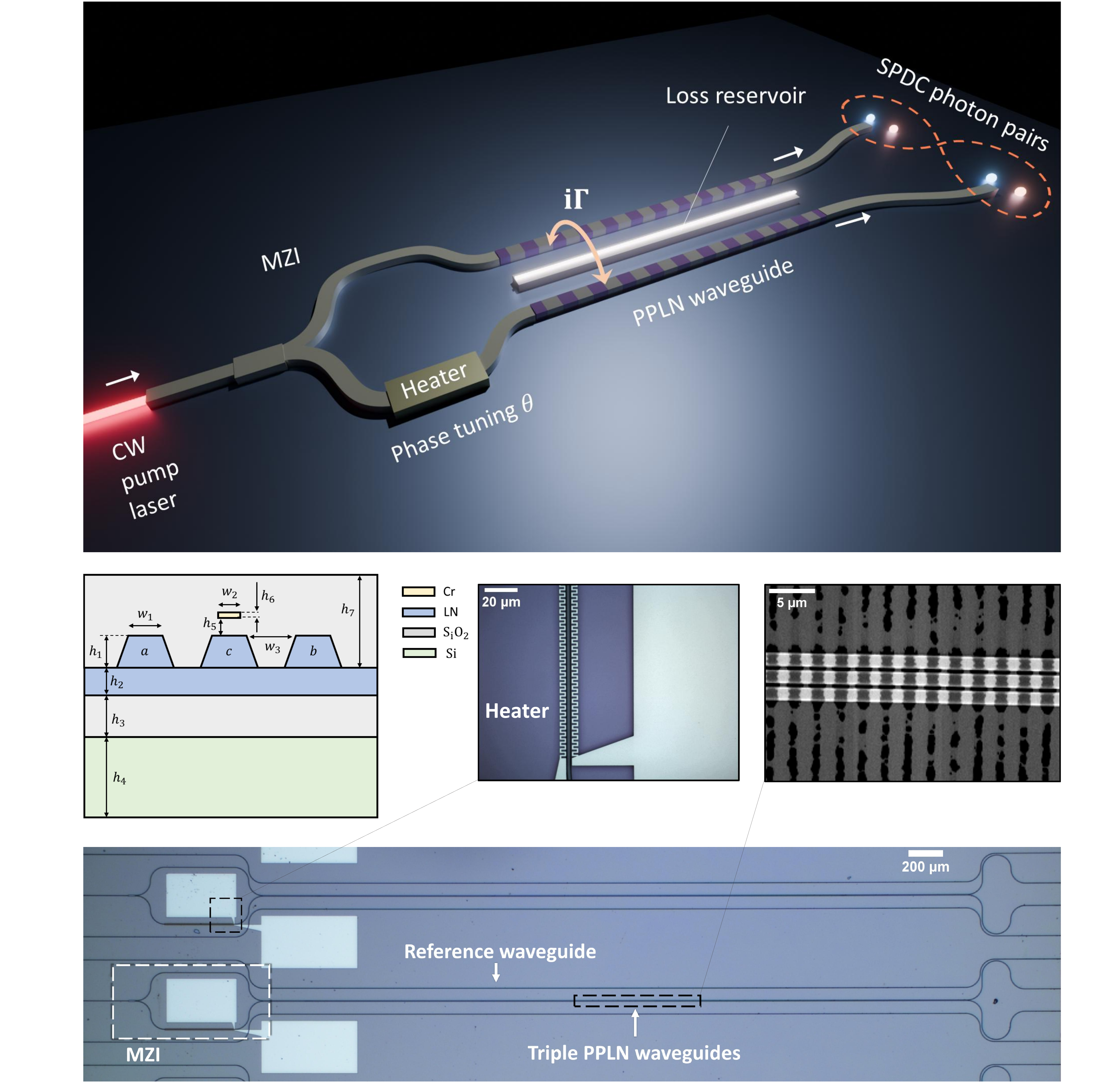}};

  \path[use as bounding box] (img.south west) rectangle (img.north east);
\node at (0.6,15.3) {\textbf{(a)}};
\node at (0.6,7.2) {\textbf{(b)}};
\node at (0.6,3.4) {\textbf{(c)}};

\end{tikzpicture}
\vspace{4pt}
\caption{\textbf{Anti-PT system realized with a dissipatively coupled triple-waveguide structure on a TFLN platform}.
(a) Chip schematic. The structure consists of three periodically poled lithium niobate (PPLN) waveguides, each with a length of 4 mm and a poling period of 3.25~$\mu$m. The system can be effectively modeled as the interaction between the modes of two waveguides $a$ and $b$, wherein the generated SPDC photon pairs undergo dissipative coupling at an effective rate $\Gamma$. A heater surrounding one arm of a Mach–Zehnder interferometer (MZI) is used to thermally tune the relative phase between the pump fields injected into waveguides $a$ and $b$ via the thermo-optic effect in lithium niobate. The loss reservoir is realized by depositing a chromium (Cr) strip on top of the middle waveguide.
(b) Cross-sectional geometry of the triple-waveguide region. Design parameters: waveguide top width $w_1=1.37~\mu$m, waveguide height $h_1=348$ nm, Cr width $w_2=250$ nm, inter-waveguide gap $w_3=600$ nm, lithium niobate slab thickness $h_2=252$ nm, SiO$_2$ undercladding thickness $h_3=2~\mu$m, Si substrate thickness $h_4=0.5$ mm, vertical separation between Cr and the middle waveguide $h_5=100$ nm, Cr thickness $h_6=20$ nm, and SiO$_2$ top-cladding thickness $h_7=1.5~\mu$m.
(c) Microscope images of the fabricated chip. The MZI is connected to the waveguides $a$ and $b$ of the triple PPLN waveguides, which are in turn connected to the two output waveguides with a separation of 127 $\mu$m at the chip facet. A reference waveguide with identical geometry is used to experimentally extract the nonlinear conversion coefficient (see Methods). Insets: a scanning electron microscope (SEM) image of the ferroelectric domain inversion in the triple PPLN waveguides, and a zoomed-in optical microscope image of the zigzag heaters surrounding one arm of the MZI.}\label{fig1}
\end{figure}
\begin{figure}[h]
\centering
\begin{tikzpicture}
\node[anchor=south west, inner sep=0] (img) at (0,0)
    {\includegraphics[width=0.8\linewidth]{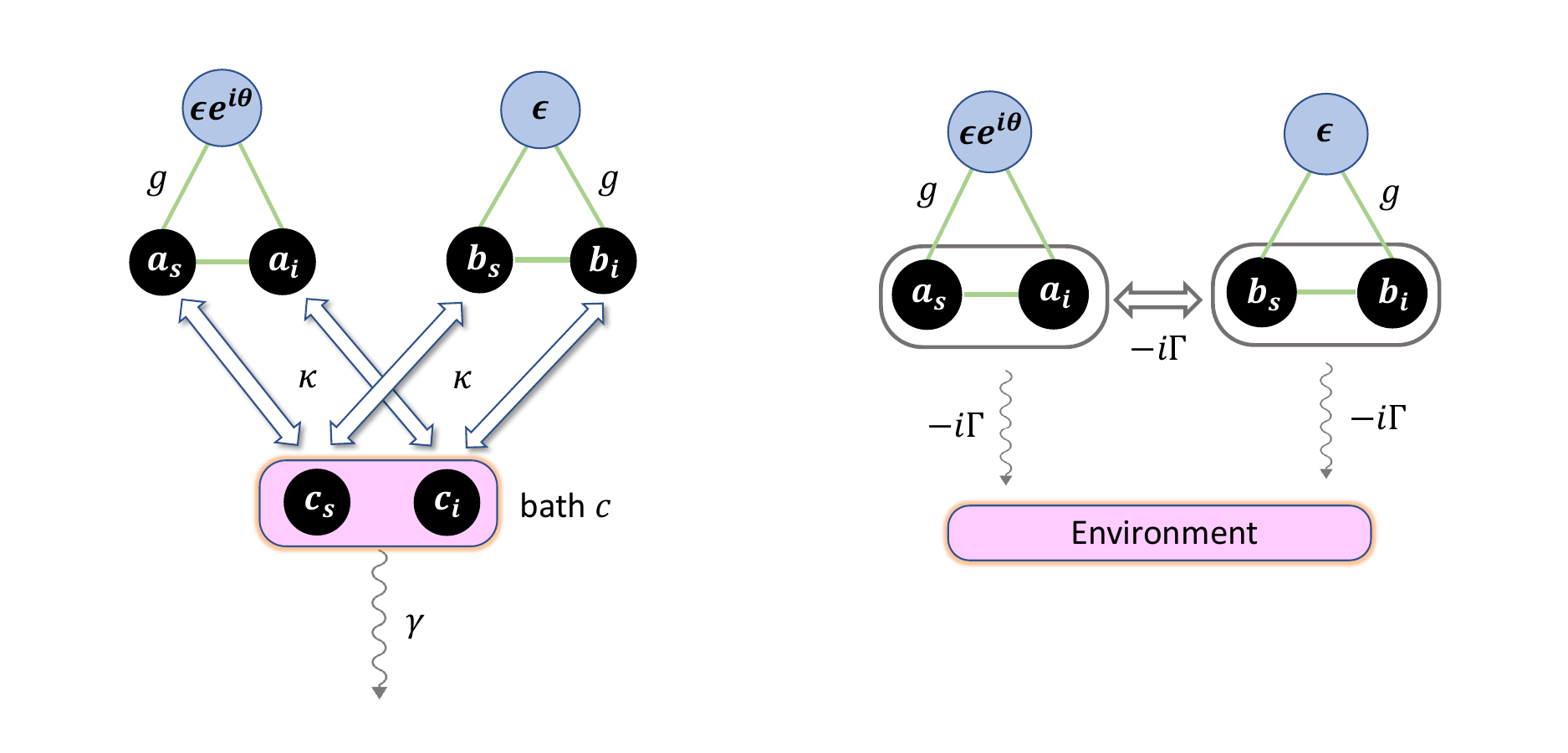}};
\node at (0.6,5.6) {\textbf{(a)}};
\node at (7.1,5.6){\textbf{(b)}};
\end{tikzpicture}
\caption{\textbf{System model and its equivalence by adiabatic elimination}. (a) Full system dynamics prior to adiabatic elimination, including the nonlinear interaction and linear coupling to a lossy auxiliary waveguide $c$, with the interaction strengths $g$ and $\kappa$, respectively.  (b) Effective Hamiltonian picture after adiabatic elimination, highlighting the induced effective interaction between waveguides $a$ and $b$, as well as the local dissipation to the environment, both characterized by the imaginary rate $-i\Gamma$. 
}\label{fig2}
\end{figure}
\noindent The chip device is fabricated on a z-cut, 600-nm TFLN wafer. As illustrated in Fig.~\ref{fig1}(a), the structure comprises three waveguides: two periodically poled lithium niobate (PPLN) waveguides (4 mm length) with propagating modes $a$ and $b$, and a central waveguide with mode $c$, engineered to exhibit a high propagation loss. The dissipation of mode $c$, characterized by a propagation loss rate $\gamma$, is introduced by a chromium (Cr) strip deposited above the middle waveguide.  A continuous-wave (CW) laser is injected into the input of the Mach–Zehnder interferometer (MZI), where a multimode interferometer (MMI) splits the input into two paths. The phase of the pump light in one of the waveguide arms is thermally tuned by a zigzag-shaped heater surrounding the waveguide. 
The cross-sectional geometry of the periodically poled waveguide region is detailed in Fig.~\ref{fig1}(b). A microscope image of the complete structure of the fabricated chip, a scanning electron microscope (SEM) image showing the periodic-poling contrast, and a magnified microscope image of the MZI heater pattern are presented in Fig.~\ref{fig1}(c).  The fabrication procedures and additional details of the geometric design are provided in the Methods section and Supplementary Information Section S1, respectively. The design of the zigzag heater follows the approach described in Ref.~\cite{heater}.

The system can be described by the Hamiltonian 
\begin{equation}\label{H_tot}
    H=H_{\rm NL}+H_{\rm L}.
\end{equation}
Here, $H_{\rm NL}$ describes the nonlinear SPDC process that generates non-degenerate signal-idler photon pairs in the two waveguides $a$ and $b$, with the corresponding modes denoted by  $a_{s,i}$ and $b_{s,i}$ ($\hbar=1$):
\begin{equation}
    H_{\rm NL}=g\epsilon(e^{i\theta}a_s^{\dagger}a^{\dagger}_i+e^{-i\theta}a_s a_i)+g\epsilon(b_s^{\dagger}b_i^{\dagger}+b_s b_i),
\end{equation}  
where $g$ is the nonlinear coupling coefficient, $\epsilon$ is the classical pump amplitude, and $\theta$ denotes the relative phase between the pumps driving the two PPLN waveguides. 
The second term in Eq.~(\ref{H_tot}) is the linear coupling term 

\begin{equation}
    H_{\rm L}=\kappa_s(a_s^{\dagger}c_s+b^{\dagger}_sc_s+\text{h.c.})+\kappa_i(a_i^{\dagger}c_i+b_i^{\dagger}c_i+\text{h.c.}),
\end{equation}
where $c_{s,i}$ are the signal and idler modes propagating in the waveguide $c$. $\kappa_{s,i}$ is the coupling strength between the modes $a_{s,i}$ and $c_{s,i}$, or between the modes $b_{s,i}$ and $c_{s,i}$. Note that we neglect the nonlinear terms $g\epsilon_c c_s^{\dagger}c_i^{\dagger} + \mathrm{h.c.}$ in $H_{\rm NL}$, assuming the pump field strength $\epsilon_c\approx0$ in waveguide $c$ due to weak coupling (see Supplementary Information Section S1). Therefore, the modes $c_{s,i}$ in $H_{\rm L}$ are not generated locally in waveguide $c$ via SPDC, but originate from linear hopping of the signal and idler modes generated in waveguide $a$ or $b$.  

The adiabatic elimination of the lossy intermediate modes $c_s$ and $c_i$ can be performed independently under the condition $\gamma\gg|\kappa_{s,i}|$. Following the adiabatic elimination procedures \cite{Anti_PT1}, and assuming $|\kappa_s|\approx|\kappa_i|=|\kappa|$, each frequency mode produces a collective jump
operator, $c_{s,i}=-\frac{i\kappa^*}{\gamma}(a_{s,i}+b_{s,i})$, and the linear part of the system is therefore effectively reduced from three modes to two, yielding a non-Hermitian effective Hamiltonian:
\begin{equation}
    H_\mathrm{L}'=-i\Gamma\sum_{\mu=s,i}(a_{\mu}^{\dagger}a_{\mu}+a_{\mu}^{\dagger}b_{\mu}+b_{\mu}^{\dagger}a_{\mu}+b_{\mu}^{\dagger}b_{\mu}),
\end{equation}
with an effective coupling strength $\Gamma=|\kappa|^2/\gamma$. The reduced linear coupling terms of the Hamiltonian possess anti-PT symmetry, satisfying $\{PT, H_\mathrm{L}'\}=0$. For $\Gamma\gg g\epsilon$, the nonlinear SPDC process described by $H_{\rm NL}$ can be treated as a perturbation to the anti-PT system. Figure~\ref{fig2} illustrates the model reduction from the three-mode description (the original) to the two-mode effective description (after the elimination). Specifically, the shared bath $c_{\mu}$ mediates the interactions between modes $a_{\mu}$ and $b_{\mu}$, resulting in an effective two-mode model characterized by dissipative coupling terms $a_{\mu}^{\dagger}b_{\mu}+b_{\mu}^{\dagger}a_{\mu}$ and local decay terms $a_{\mu}^{\dagger}a_{\mu}+b_{\mu}^{\dagger}b_{\mu}$.

To simulate the system dynamics, we use two approaches: the Lindblad master equation approach and the non-Hermitian effective Hamiltonian approach. In the former, the spatial evolution of the density operator $\rho$ is given by
\begin{equation}\label{master}
    \frac{d\rho}{dz}=-i[H_{\rm NL},\rho]+2\Gamma \sum_{\mu=s,i}\mathcal{D}[a_{\mu}+b_{\mu}]\rho.
    \end{equation}
Here, the dissipator is defined in the standard Lindblad form, $\mathcal{D}[O]\rho=O\rho O^{\dagger}-\frac{1}{2}\{O^{\dagger}O,\rho\}$. The first term (quantum jumps) of the dissipator describes the state update induced by continuous monitoring from the environment \cite{QP4,QP6} and the second term describes photon loss through the collective decay modes $c_{\mu}$. In the second approach, governed by the non-Hermitian effective Hamiltonian, the spatial Schrödinger equation is given by 
\begin{equation}\label{Sch}
    i\partial_z|\Psi(z)\rangle=H_{\rm eff}|\Psi(z)\rangle,
\end{equation}
where the total effective Hamiltonian reads $H_\mathrm{eff}=H_\mathrm{NL}+H_\mathrm{L}'$. The state of the system is expanded in the Fock basis as
\begin{equation}
    |\Psi(z)\rangle=\sum_{m_s,m_i,n_s,n_i} C_{m_s,m_i,n_s,n_i}(z)|m_s,m_i;n_s,n_i\rangle,
\end{equation}
where $m_{\mu}$ and $n_{\mu}$  denote the photon numbers for modes $\mu\in\{s,i\}$ in waveguides $a$ and $b$, respectively, and $C_{m_s,m_i,n_s,n_i}(z)$ are the probability amplitudes of the corresponding Fock states. 
 The Lindblad master equation can be rearranged into the form
\begin{equation}
    \frac{d\rho}{dz}=-i(H_{\rm eff}\rho-\rho H_{\rm eff}^{\dagger})+2\gamma\sum_{\mu=s,i}c_{\mu}\rho c^{\dagger}_{\mu}.
\end{equation}
The equation described by the first term alone represents the von Neumann evolution, which is equivalent to Eq.~(\ref{Sch}). 
This reformulation demonstrates that the fundamental difference between the two approaches lies in the quantum jump terms, $\mathcal{J}\rho = 2\gamma\sum_{\mu} c_{\mu}\rho c^{\dagger}_{\mu}=2\Gamma\sum_{\mu}(a_{\mu}+b_{\mu})\rho(a_{\mu}^{\dagger}+b_{\mu}^{\dagger})$, which are neglected in the effective Hamiltonian method.


To examine the effect of dissipative coupling, we compare the above anti-PT linked system to a coherent, evanescently coupled system. Specifically, we consider two PPLN waveguides coupled coherently, described by the Hamiltonian
\begin{equation}\label{coherent}
    H_{\rm c}=H_{\rm NL}+\Gamma\sum_{\mu=s,i}(a_{\mu}^{\dagger}b_{\mu}+a_{\mu}b_{\mu}^{\dagger}).
\end{equation}

We are interested in comparing the four-photon quantum interference of the two signal-idler photon pairs in the systems described above. The normalized fourth-order correlation among the four SPDC photons is defined as
\begin{equation}\label{eq}
    g^{(4)}=\frac{\langle a_s^{\dagger}a_i^{\dagger}b_s^{\dagger}b_i^{\dagger}b_ib_sa_ia_s\rangle}{\langle a^{\dagger}_sa_s\rangle\langle a^{\dagger}_ia_i\rangle\langle b^{\dagger}_sb_s\rangle\langle b^{\dagger}_ib_i\rangle},
\end{equation}
which is normalized with respect to the mean photon numbers in each mode, e.g., $\langle n_{a_s}\rangle=\langle a_s^{\dagger} a_s\rangle$. Although the standard fourth-order correlation $g^{(4)}$ effectively distinguishes four-photon coincidences from the accidental overlap of four independent single photons, it does not provide information to distinguish nontrivial four-photon interference from the trivial accidental coincidence of two independent signal-idler pairs, i.e., even when the two photon pairs are independent, $g^{(4)}=\langle a_s^{\dagger}a_i^{\dagger}a_ia_s\rangle \langle b_s^{\dagger}b_i^{\dagger}b_ib_s\rangle/(\langle a^{\dagger}_sa_s\rangle\langle a^{\dagger}_ia_i\rangle\langle b^{\dagger}_sb_s\rangle\langle b^{\dagger}_ib_i\rangle)=g^{(2)}_a g^{(2)}_b$ still yields a substantial correlation signature. Here, $g^{(2)}_a=\langle a_s^{\dagger}a_i^{\dagger}a_ia_s\rangle/(\langle a^{\dagger}_sa_s\rangle\langle a^{\dagger}_ia_i\rangle)$ and $g^{(2)}_b=\langle b_s^{\dagger}b_i^{\dagger}b_ib_s\rangle/(\langle b^{\dagger}_sb_s\rangle\langle b^{\dagger}_ib_i\rangle)$ represent the two-photon correlation of photon pairs from waveguides $a$ and $b$, respectively. Hence, to characterize the correlation between the photon pairs, we define the four-photon inter-pair correlation ratio 
\begin{equation}\label{eqnR}
    \mathcal{R}^{(4)}=\frac{\langle a_s^{\dagger}a_i^{\dagger}b_s^{\dagger}b_i^{\dagger}b_ib_sa_ia_s\rangle}{\langle a_s^{\dagger}a_i^{\dagger}a_ia_s\rangle\langle b_s^{\dagger}b_i^{\dagger}b_ib_s\rangle},
\end{equation}
which is simply $g^{(4)}/g_{a}^{(2)}g_{b}^{(2)}$. We emphasize that the higher-order correlations studied here do not imply irreducible higher-order correlations in the sense of non-Gaussian connected moments. As shown in Supplementary Section S2, the state remains Gaussian, such that higher-order field moments can be reduced to combinations of second-order moments according to Wick’s theorem \cite{Wick,wolf,Gaussian}. Here, by nontrivial multiphoton interference we instead refer to phase-sensitive correlations between distinct photon-pair generation pathways, quantified by $\mathcal{R}^{(4)}$, in the dissipative anti-PT system. Such Gaussian multiphoton correlations, although reducible, are well established in coherently coupled Hermitian systems \cite{four_photon2,Gaussian_multi,Gaussian_multi_2,Gaussian_multi_3} and remain physically significant resources  for Gaussian boson sampling, including graph optimization and molecular spectroscopy, as well as for broader Gaussian quantum information processing and sensing \cite{Gaussian_boson_sampling,Gaussian_app1,Gaussian_app2,Gaussian,Gaussian_app3,Gaussian_app4,Gaussian_app5}.

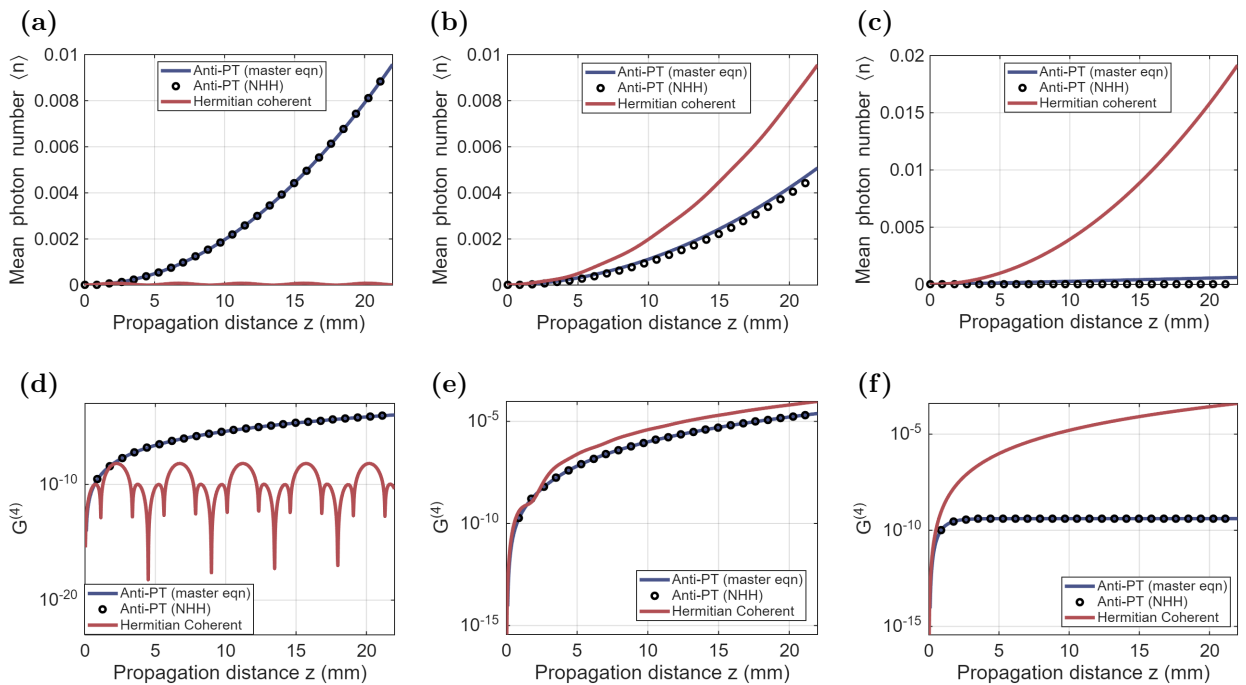
\begin{figure}[h]
\centering
\begin{tikzpicture}
\node[anchor=south west, inner sep=0] (img) at (0,0)
    {\includegraphics[width=\linewidth]{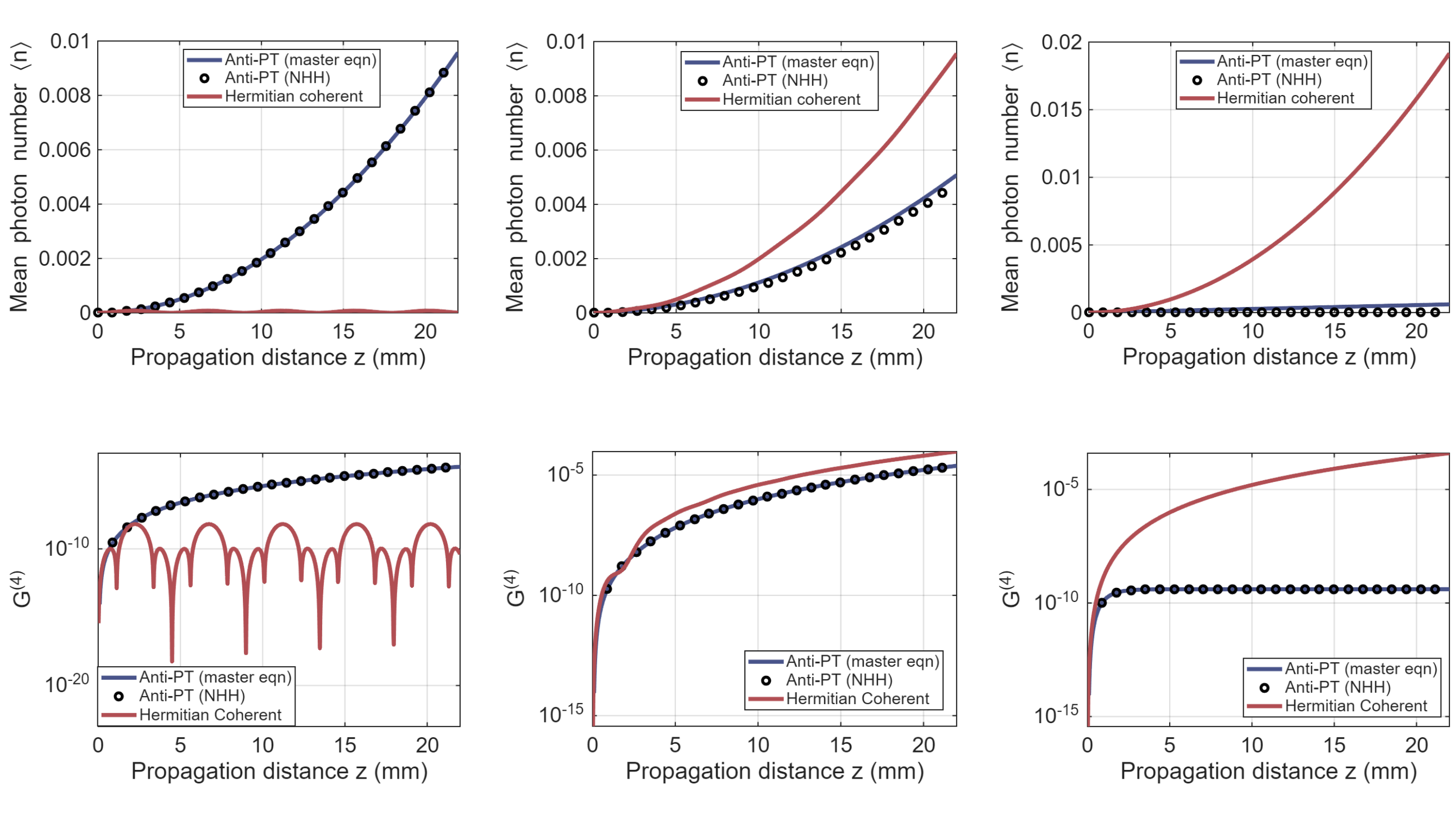}};

  \path[use as bounding box] (img.south west) rectangle (img.north east);
\node at (0.5,9.1) {\textbf{(a)}};
\node at (5.9,9.1) {\textbf{(b)}};
\node at (11.5,9.1) {\textbf{(c)}};
\node at (0.5,4.3) {\textbf{(d)}};
\node at (5.9,4.3) {\textbf{(e)}};
\node at (11.5,4.3) {\textbf{(f)}};
\end{tikzpicture}
\caption{\textbf{Numerical results of system dynamics across pump relative phases}. Evolutions of (a–c) the mean photon number $\langle n\rangle$ and (d–f) the unnormalized four-photon correlation 
$G^{(4)}$ in the anti-PT system, comparing the Lindblad master-equation model with the non-Hermitian effective-Hamiltonian (NHH) model, and with the Hermitian coherent-coupling reference system. Results are shown for pump relative phases (a, d) $\theta=0$, (b, e) $\theta=\pi/2$, and (c, f) $\theta=\pi$. The master-equation and NHH approaches yield identical $G^{(4)}$
 dynamics, while 
$\langle n\rangle$ exhibits discrepancies between the two approaches at $\theta\neq 0$ because quantum-jump terms are included in the master equation but absent in the NHH description. In the Hermitian system, the additional propagation phases introduced by coherent coupling disrupt phase matching in the PPLN waveguides, inducing oscillatory behavior in the evolution. In contrast, dissipative coupling in the anti-PT system leaves the phase matching unaffected.} \label{fig3}
\end{figure}

Figure~\ref{fig3} presents numerical simulation results obtained by solving the equations for the anti-PT system using both the effective Hamiltonian and Lindblad master equation approaches, alongside results for the reference coherent system. We utilize the parameters $g\epsilon=6.92~\text{m}^{-1}$ and $\Gamma=7.22~\text{cm}^{-1}$, applying a Fock basis truncation of $m_{\mu}, n_{\mu} \leq N_{\text{max}}$ where $N_{\text{max}}=10$. We study the mean photon number dynamics of the above three cases under different phases. Meanwhile, to visualize the underlying four-photon production dynamics itself, which is directly proportional to experimental observations of fourfold coincidence events, we also plot the unnormalized four-photon correlation function $G^{(4)}(z)=\langle a_s^{\dagger}a_i^{\dagger}b_s^{\dagger}b_i^{\dagger}b_ib_sa_ia_s\rangle$ versus the propagation distance $z$. For 
$\theta=0$, as shown in Fig.~\ref{fig3}(a), the Hermitian coherent system exhibits oscillatory mean photon-number dynamics, which in turn leads to the corresponding fluctuations in $G^{(4)}(z)$ observed in Fig.~\ref{fig3}(d). To elucidate the underlying physics, we perform a unitary transformation from the physical waveguide modes $(a_{\mu},b_{\mu})$ to the bright and dark supermodes $(B_{\mu},D_{\mu})$. The bright and dark modes are defined as $ B_{\mu}=\frac{a_{\mu}+b_{\mu}}{\sqrt{2}}$, $D_{\mu}=\frac{a_{\mu}-b_{\mu}}{\sqrt{2}}$
for $\mu\in \{s,i\}$. The normalization factor $1/\sqrt{2}$ ensures that this transformation is canonical, preserving the bosonic commutation relations, i.e.,  $[B,B^{\dagger}]=1$ and $[D,D^{\dagger}]=1$. In the $B$/$D$ basis, the nonlinear SPDC Hamiltonian becomes 
\begin{equation}\label{NL}
    H_{\rm NL}(B,D)=\frac{g\epsilon}{2} \bigl[(e^{i\theta}+1)(B_{s}^{\dagger}B_i^{\dagger}+D_s^{\dagger}D_i^{\dagger})+(e^{i\theta}-1)(B_s^{\dagger}D_i^{\dagger}+D_s^{\dagger}B_i^{\dagger})\bigr]+\text{h.c.}
\end{equation}
The oscillations at $\theta=0$ arise from the extra propagation phase accumulated during the coherent evolution ($B_{\mu}\propto e^{-i\Gamma z}$, $D_{\mu}\propto e^{i\Gamma z}$), which renders the phase-matching condition unfulfilled. Because the bright and dark modes acquire opposite extra phases, at $\theta=\pi$—where the pair-creation (or annihilation) process involves one bright and one dark mode, i.e., $B_s^{\dagger}D_i^{\dagger}+D_s^{\dagger}B_i^{\dagger}+\mathrm{h.c}.$—the effects of the extra propagation phases on the phase matching cancel each other, resulting in monotonic growth of the dynamics (Fig.~\ref{fig3}(c) and \ref{fig3}(f)). In the anti-PT symmetry system, phase matching remains undisturbed because the bright modes are governed by dissipative evolution ($B_{\mu}\propto e^{-2\Gamma z}$) and the dark modes remain lossless, under the condition that the decay rate dominates the nonlinear coupling coefficient ($\Gamma\gg g\epsilon$). As a result, both the mean photon number and $G^{(4)}$ accumulate monotonically during propagation, regardless of the relative phase. The origin of the extra propagation phase is evident from the equations of motion for arbitrary $\theta$, provided in Supplementary Information Section S2, where analytical solutions are given for $\theta=0$ and $\pi$. For the intermediate phase $\theta=\pi/2$, the dynamics of the simulation are shown in Fig.~\ref{fig3}(b) and \ref{fig3}(e). 

Although the dynamics of $G^{(4)}(z)$ remain identical for the non-Hermitian Hamiltonian and master equation approaches in the anti-PT system, there is a discrepancy between the two approaches in the mean photon number for non-zero relative phases. This discrepancy is attributed to the quantum jump terms. Consequently, the master equation approach is required to provide the fully accurate dynamics of the normalized correlation $g^{(4)}(z)$, which depends strongly on the brightness (mean photon number). In contrast, for the inter-pair correlation ratio $\mathcal{R}^{(4)}$, the non-Hermitian Hamiltonian approach remains a sufficient approximation. Further detailed analysis is provided in Supplementary Information Section S3. 


The robustness of the anti-PT system can be further analyzed in the bright and dark states. Since the state of the system can be described in the bases of these collective excitations in the Fock space: $|n_{B_s},n_{B_i};n_{D_s},n_{D_i}\rangle$, the bright state can be described as $|n_{B_s},n_{B_i};0,0\rangle=(B^{\dagger}_{s})^{n_{B_s}}(B^{\dagger}_{i})^{n_{B_i}}|0,0;0,0\rangle\\/\sqrt{n_{B_s}!n_{B_i}!}$, and the dark state as $|0,0;n_{D_s},n_{D_i}\rangle=(D^{\dagger}_{s})^{n_{D_s}}(D^{\dagger}_{i})^{n_{D_i}}|0,0;0,0\rangle/\sqrt{n_{D_s}!n_{D_i}!}$. 
The master equation in the $B$/$D$ basis can be reformulated as
\begin{equation}\label{BDmaster}
    \frac{d\rho}{dz}=-i\bigl[H_{\rm NL}(B,D),\rho\bigr]+4\Gamma\sum_{\mu=s,i}\mathcal{D}[B_{\mu}]\rho.
\end{equation}
Notably, the dark mode operators $D_{\mu}$ vanish entirely from the dissipative term, revealing the selective nature of the loss channel. The loss exclusively acts on the bright state, whereas the dark state remains transparent to dissipation, residing in a Decoherence-Free Subspace (DFS) \cite{DFS}. Note that this is an exact DFS of the dissipator, but not a strict DFS of the full Liouvillian. The collective dissipation continuously removes bright components, leading to dissipative purification toward the dark manifold as a global attractor.

The equations of motion for the mean photon numbers in each manifold, calculated as $\langle n_B\rangle=\langle B^{\dagger}B\rangle$ and $\langle n_D\rangle=\langle D^{\dagger}D\rangle$, are given by
\begin{align}
\frac{d\langle n_{B_s}\rangle}{dz}
&= -4\Gamma\langle n_{B_s}\rangle
   -2\operatorname{Im}\!\left[
      \Lambda^*_{\rm co}\langle B_sB_i\rangle
      +\Lambda^*_{\rm x}\langle B_sD_i\rangle
   \right],
\label{eqnBs}\\
\frac{d\langle n_{B_i}\rangle}{dz}
&= -4\Gamma\langle n_{B_i}\rangle
   -2\operatorname{Im}\!\left[
      \Lambda^*_{\rm co}\langle B_sB_i\rangle
      +\Lambda^*_{\rm x}\langle D_sB_i\rangle
   \right],
\label{eqnBi}\\
\frac{d\langle n_{D_s}\rangle}{dz}
&= -2\operatorname{Im}\!\left[
      \Lambda^*_{\rm co}\langle D_sD_i\rangle
      +\Lambda^*_{\rm x}\langle D_sB_i\rangle
   \right],
\label{eqnDs}\\
\frac{d\langle n_{D_i}\rangle}{dz}
&= -2\operatorname{Im}\!\left[
      \Lambda^*_{\rm co}\langle D_sD_i\rangle
      +\Lambda^*_{\rm x}\langle B_sD_i\rangle
   \right].
\label{eqnDi}
\end{align}
Here, $\Lambda_{\rm co}=g\epsilon e^{i\theta/2}\cos(\theta/2)$ and $\Lambda_{\rm x}=ig\epsilon e^{i\theta/2}\sin(\theta/2)$ are the co-generation and cross-generation amplitudes, respectively. As indicated by Eqs.~\eqref{eqnBs}--\eqref{eqnDi}, the bright photon experiences dissipation at a rate of $4\Gamma$, while the dark photon remains protected and therefore exhibits a significantly longer lifetime, given the condition $\Gamma\gg g\epsilon$. 

The dominance of the effective linear coupling rate $\Gamma$ causes the bright and dark states in the quantum anti-PT system to inherit the hallmark behavior of classical anti-PT systems (without nonlinear processes involved) within the symmetry-unbroken phase, where purely imaginary eigenvalues give rise to distinct attenuation rates for the two eigenstates of the anti-PT effective Hamiltonian \cite{Anti_PT1,Anti_PT2}. From a practical perspective, the effect is essential for creating error-resilient and DFS-based quantum light sources \cite{DFS}. The dissipative structure provides a robust anti-PT protection mechanism against disorder: while fabrication imperfections typically break the destructive interference required for a perfect dark state, any resulting scattering into the bright subspace is immediately suppressed by the Cr strip, thereby biasing the dynamics toward the protected dark manifold.

\section{Experiment}\label{sec3}

\begin{figure}[h]
\centering
\includegraphics[width=\textwidth]{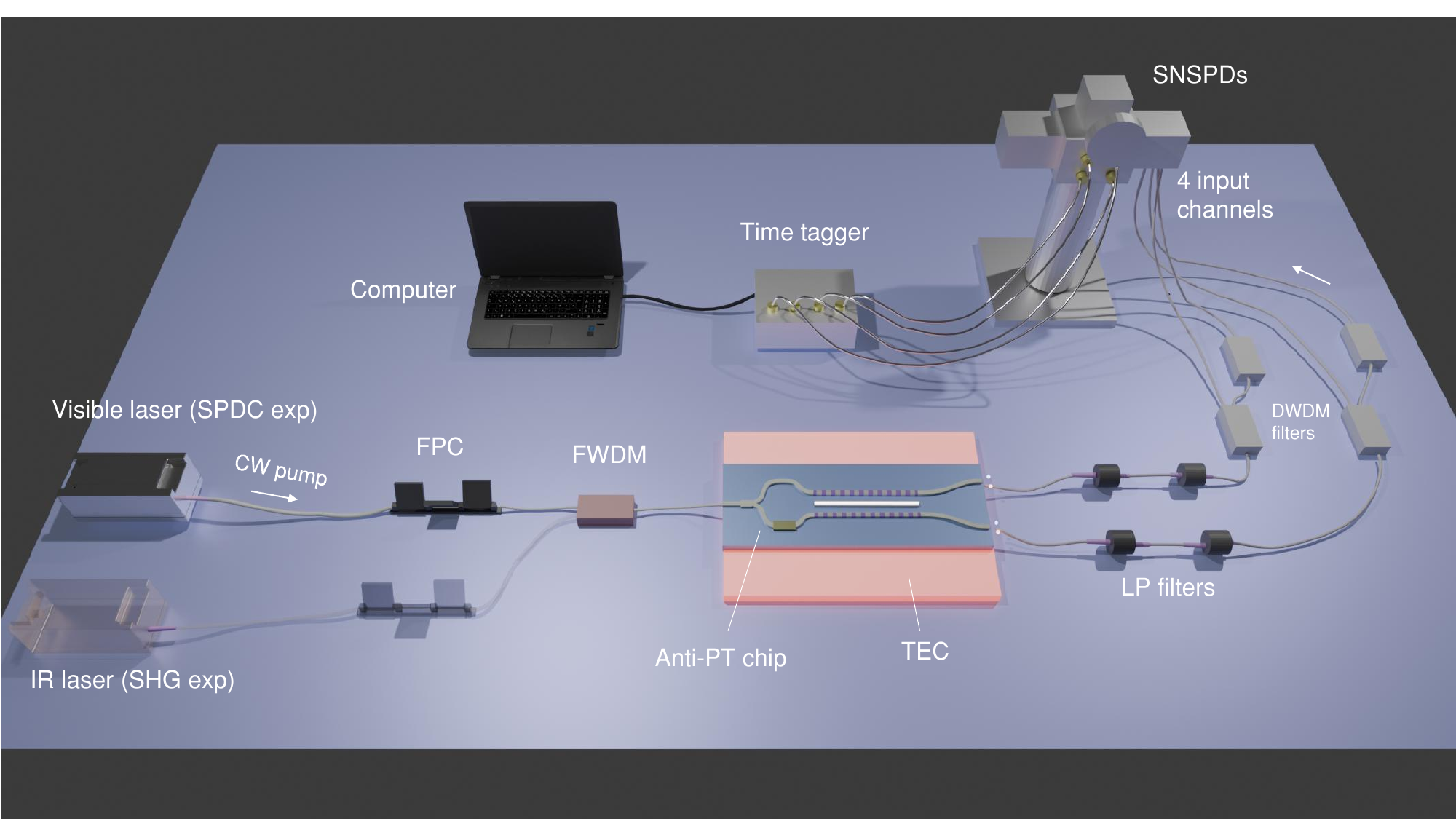}
\vspace{0.1pt}
\caption{\textbf{Experimental setup for SHG characterization and SPDC photon-correlation measurement}. For the SHG experiment, a continuous-wave (CW) input laser sweeping from 1500 to 1630 nm is polarization-controlled using a fiber polarization controller (FPC) and routed to the device through a fused wavelength division multiplexer (FWDM), while the generated second-harmonic power is characterized at the chip outputs. For the central SPDC experiment, a CW pump laser at 775.4 nm, with its polarization tuned by an FPC, is launched into the chip through the FWDM. The generated photon pairs from the two output paths are coupled into lensed fibers, while the pump light in each path is filtered out by two long-pass (LP) filters with a total suppression of 140 dB. The signal-idler photon pairs are demultiplexed by two dense wavelength division
multiplexing (DWDM) filters in each path, yielding four channels that are routed to the superconducting nanowire single-photon detectors (SNSPDs). The detector outputs are recorded by a time tagger for four-photon-correlation analysis. The chip temperature is maintained at 69 \textdegree C by a thermoelectric cooler (TEC) with a stability of $\pm 3$ mK.}\label{setup}
\end{figure}

The experimental setup is shown in Fig.~\ref{setup}. We initially characterize the quasi-phase-matching (QPM) condition of the PPLN waveguides by measuring the second-harmonic generation (SHG) power as a function of the fundamental wavelength. Guided by the QPM parameters determined from this characterization, we perform the SPDC experiment by coupling a 775.4-nm CW pump laser into the anti-PT sample. The SPDC photon pairs generated in the waveguides undergo dissipative coupling, and four-photon correlations are recorded using superconducting nanowire single-photon detectors (SNSPDs) and a time-tagger module. Details of the experimental setup and procedures are provided in the Methods section.
\begin{figure}[h]
\centering
\begin{tikzpicture}
\node[anchor=south west, inner sep=0] (img) at (0,0)
    {\includegraphics[width=0.9\linewidth]{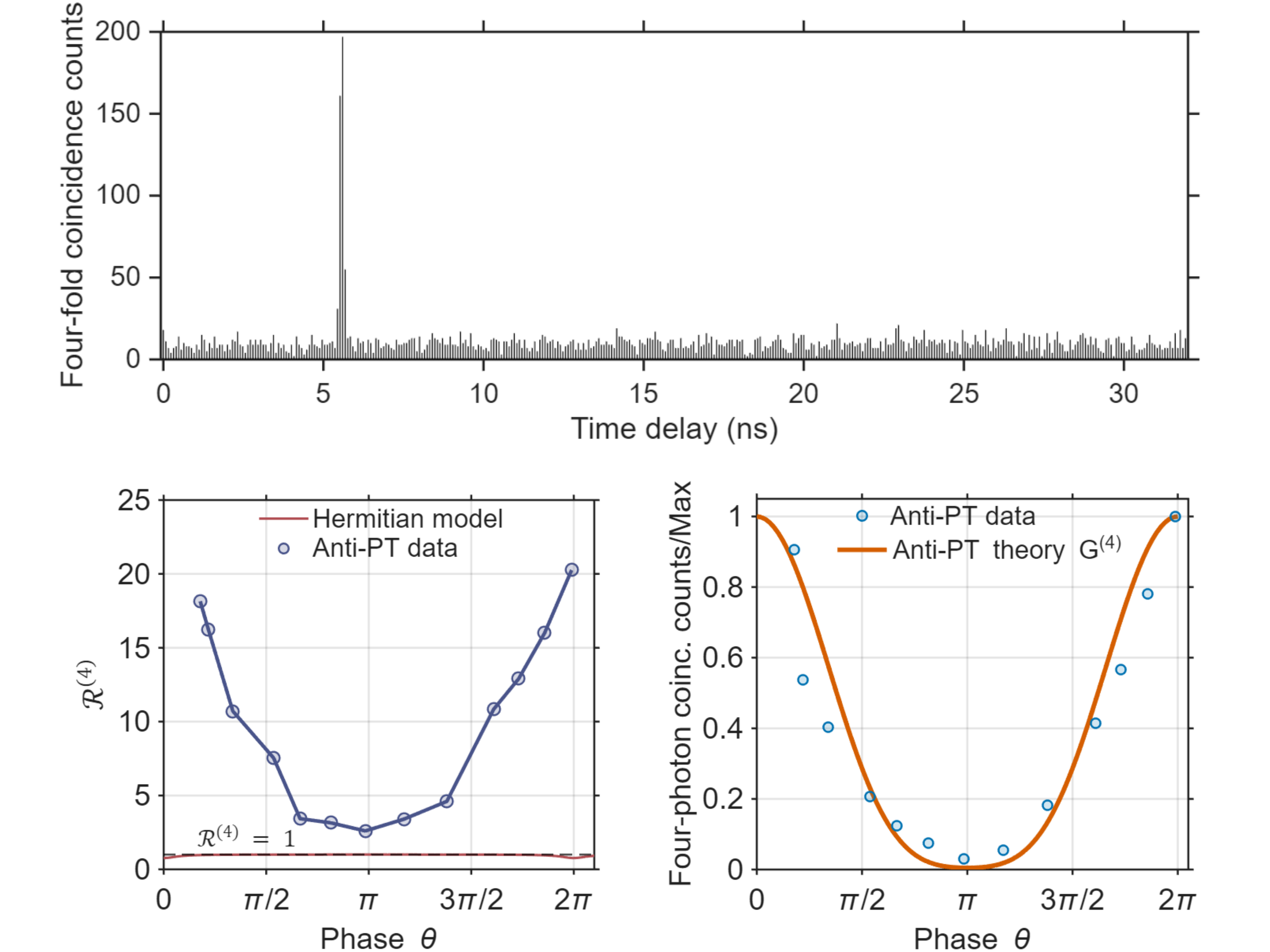}};

  \path[use as bounding box] (img.south west) rectangle (img.north east);
\node at (1,11.4) {\textbf{(a)}};
\node at (1,5.5) {\textbf{(b)}};
\node at (7.5,5.5) {\textbf{(c)}};

\end{tikzpicture}
\caption{\textbf{Four-photon correlation measurements}. (a)  Representative four-photon correlation histogram of time delays between two virtual channels defined in the time-tagger module, showing the four-photon coincidences $C_{a_s,a_i;b_s,b_i}$. (b) Phase tuning of four-photon inter-pair correlation $\mathcal{R}^{(4)}$. The $\mathcal R^{(4)}$ is determined experimentally from the fourfold coincidence-to-accidental ratio, CAR$^{(4)}=C^{(4)}_{\rm coin}/C^{(4)}_{\rm acc}$, where $C^{(4)}_{\rm coin}$ is obtained by summing the coincidence counts within two time bins (each 80 ps) of the correlation histogram in (a), and $C^{(4)}_{\rm acc}$ is computed from the average accidental counts summed over the same two-bin window. The reference level $\mathcal R^{(4)}=1$ (dashed line) indicates that the signal–idler pairs $(a_s,a_i)$ and $(b_s,b_i)$ are statistically independent of each other. The numerical prediction for a Hermitian coherent-coupling system with the same propagation length (4 mm) as the experimental anti-PT chip is shown for comparison. (c) Phase dependence of the four-photon coincidences, compared with the theoretical prediction for the unnormalized fourth-order correlation $G^{(4)}$. Experimental and theoretical results are each normalized to their respective peak values to facilitate comparison of the fringe shape and periodicity.}\label{four_data}
\end{figure}
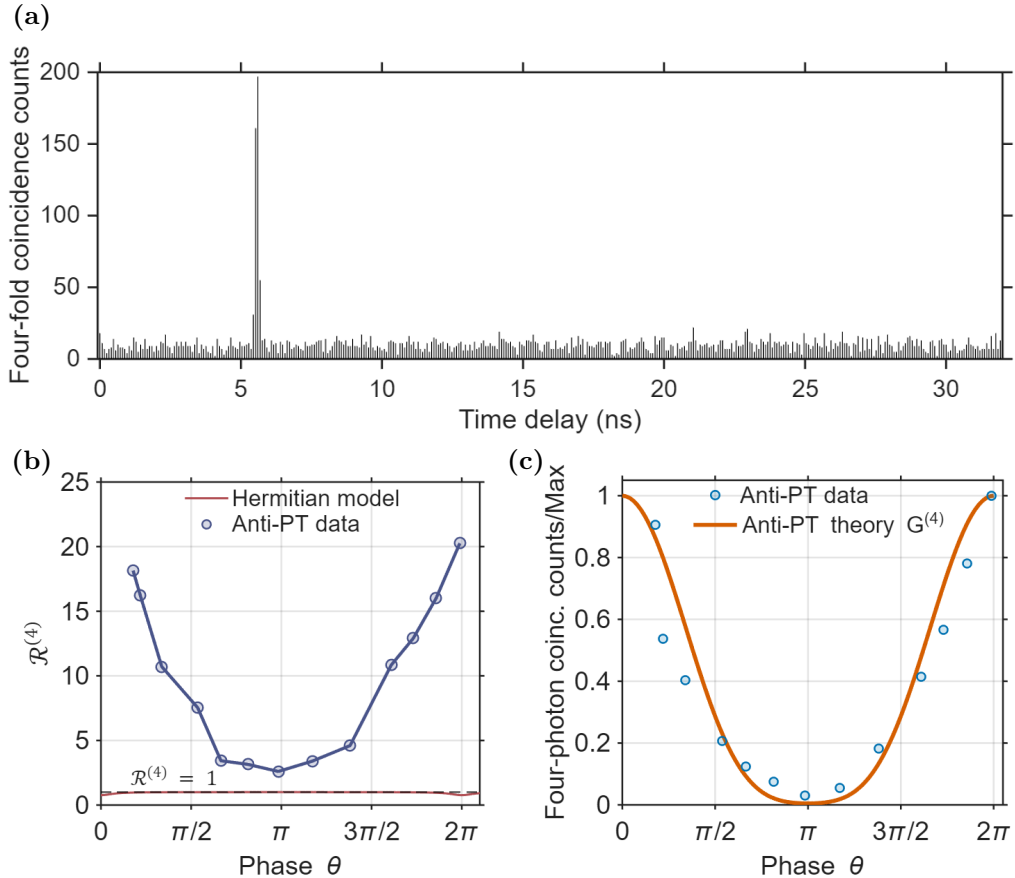
\begin{figure}[h]
\centering
\begin{tikzpicture}
\node[anchor=south west, inner sep=0] (img) at (0,0)
    {\includegraphics[width=0.9\linewidth]{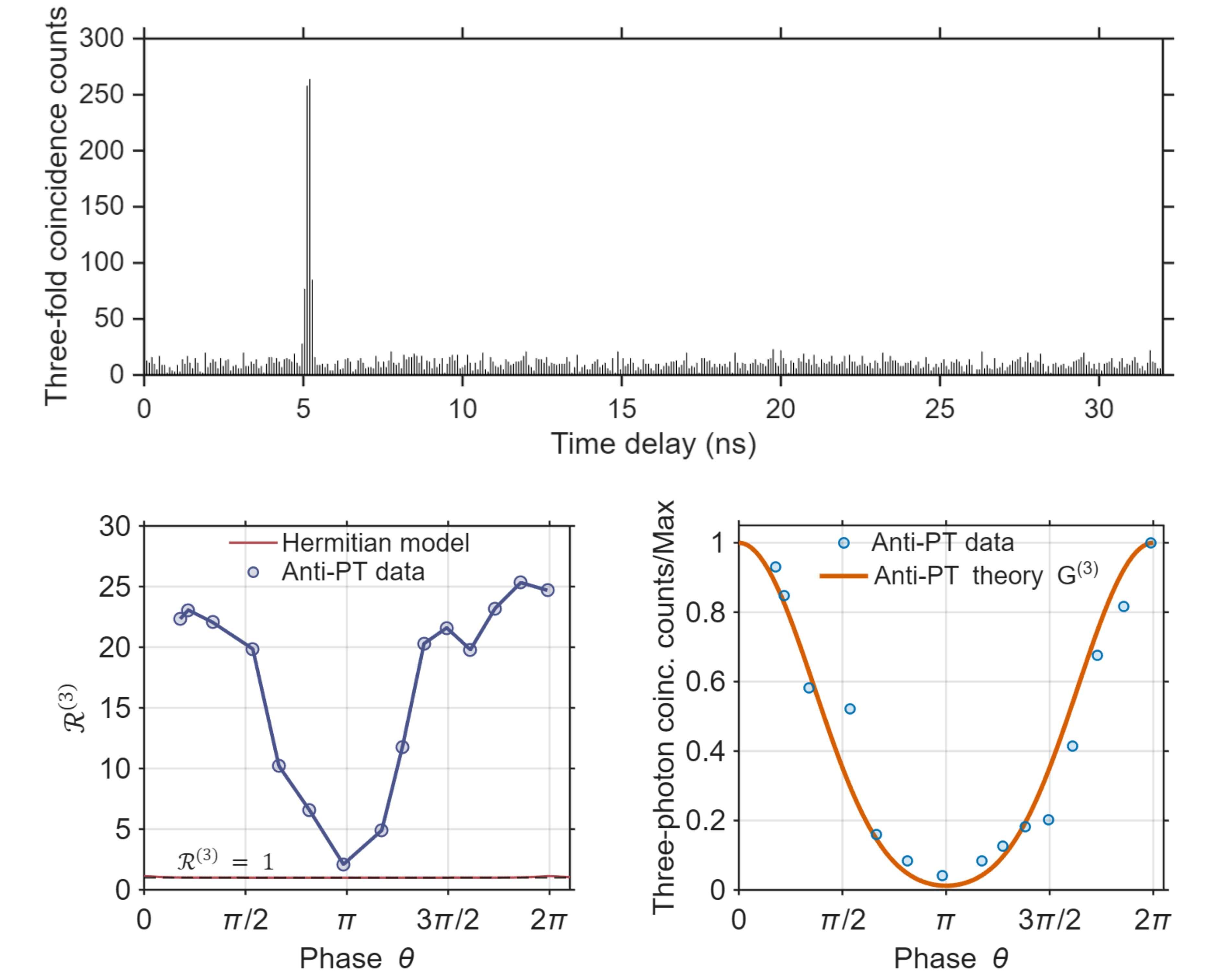}};

  \path[use as bounding box] (img.south west) rectangle (img.north east);
\node at (0.9,11.8) {\textbf{(a)}};
\node at (0.9,5.7) {\textbf{(b)}};
\node at (7.5,5.7) {\textbf{(c)}};

\end{tikzpicture}
\caption{\textbf{Three-photon correlation measurements}. (a) Representative histogram of time delays  between the virtual channel defined by the signal-idler pair $(a_s,a_i)$ and the single mode $b_i$, showing three-photon coincidences $C_{a_s,a_i;b_i}$. (b) Phase dependence of three-photon inter-pair correlation $\mathcal R^{(3)}$. The reference level $\mathcal R^{(3)}=1$ (dashed line) indicates that the signal–idler pair $(a_s,a_i)$ and the mode $b_i$ are statistically independent of each other. The numerical simulation for the Hermitian system is shown for comparison. (c) Phase dependence of the three-photon coincidences, compared with the theoretical prediction for the unnormalized third-order correlation $G^{(3)}$. Experimental and theoretical results are normalized to their peak values.}\label{three_data}
\end{figure}
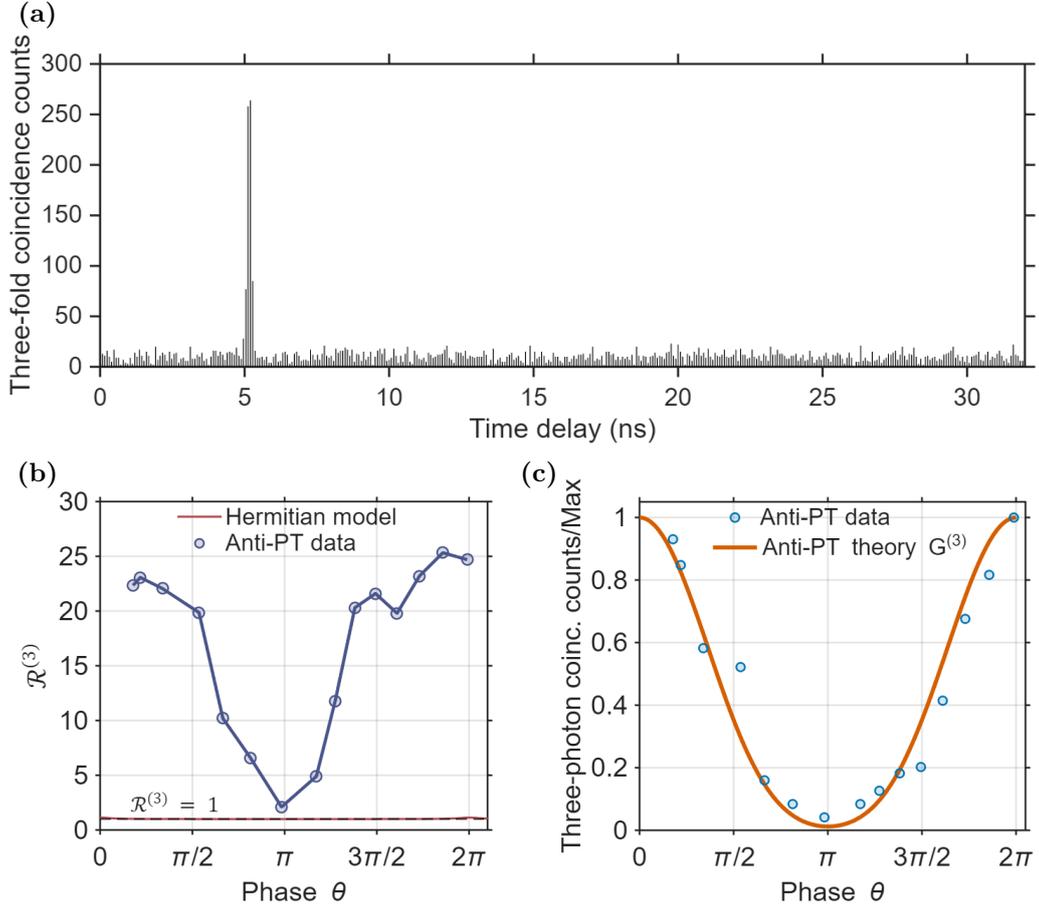
\begin{figure}[h]
\centering
\begin{tikzpicture}
\node[anchor=south west, inner sep=0] (img) at (0,0)
    {\includegraphics[width=0.95\linewidth]{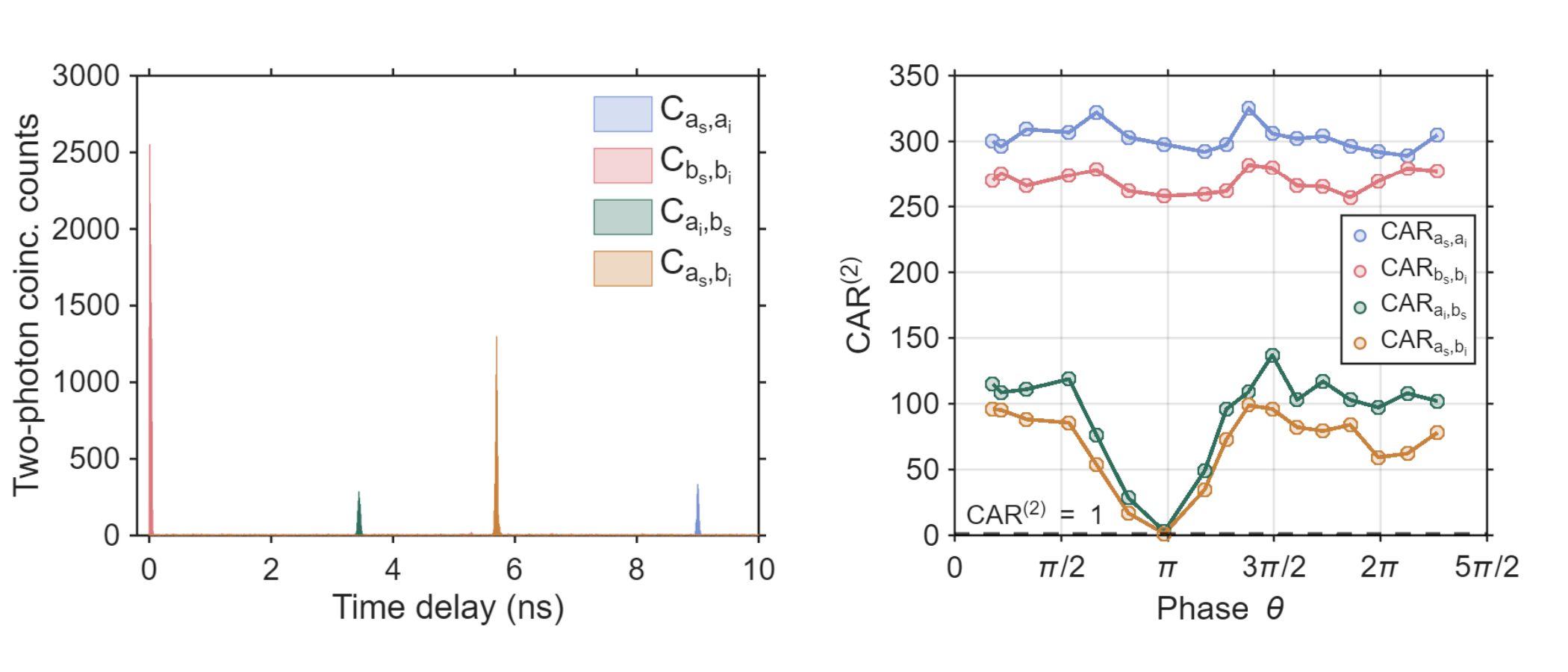}};

  \path[use as bounding box] (img.south west) rectangle (img.north east);
\node at (0.7,6.4) {\textbf{(a)}};
\node at (8.5,6.4) {\textbf{(b)}};

\end{tikzpicture}
\caption{\textbf{Two-photon correlation measurements}. (a) Histograms of time delays between two single-mode channels, showing the intra-waveguide coincidences $C_{a_s,a_i}$ and $C_{b_s,b_i}$, as well as the inter-waveguide coincidences $C_{a_i,b_s}$ and $C_{a_s,b_i}$. The differences in peak heights result from unequal propagation losses across the channels. (b) Phase dependence of two-photon coincidence-to-accidental ratios. The reference level CAR$^{(2)}=1$ (dashed line) indicates uncorrelated signal and idler photons. Clear phase dependence is observed only in the inter-waveguide CAR values, CAR$_{a_s,b_i}$ and CAR$_{a_i,b_s}$, due to the phase dependence of pair generation $a_s^{\dagger}b_i^{\dagger}$ and $a_i^{\dagger}b_s^{\dagger}$, caused by the selective action of loss on the bright and dark modes during the dissipative evolution.}\label{two_data}
\end{figure}

To measure four-photon coincidences, we define in the time-tagger module a virtual channel 1 from the twofold coincidences between the physical channels $a_s$ and $a_i$, and a virtual channel 2 from the twofold coincidences between $b_s$ and $b_i$. The four-photon correlation is then determined by measuring the coincidences between the two virtual channels, which we denote as $C_{a_s,a_i;b_s,b_i}$. Figure~\ref{four_data}(a) shows a representative four-photon correlation histogram of time delays between the two virtual channels. Next, the inter-pair correlation ratio $\mathcal{R}^{(4)}$ is calculated to confirm that the observed four-photon coincidences arise from nontrivial four-photon interference rather than the accidental coincidence of two signal-idler pairs. Based on Eq.~(\ref{eqnR}),
if the photon pairs from channels ($a_s,a_i$) and ($b_s,b_i)$ are independent of each other, the normalized four-photon correlation factorizes as $g^{(4)}=g^{(2)}_{a}\cdot g_{b}^{(2)}$, yielding a baseline of $\mathcal{R}^{(4)}=1$. This metric is equivalent to the four-photon coincidence-to-accidental ratio (CAR$^{(4)}=C_{\text{coin}}^{(4)}/C_{\text{acc}}^{(4)}$) obtained from Fig.~\ref{four_data}(a), where $C^{(4)}_\mathrm{coin}$ denotes the coincidence counts and $C^{(4)}_\mathrm{acc}$ denotes the average accidental counts, both summed over two time-delay bins in the correlation histogram. Figure~\ref{four_data}(b) illustrates the experimental results of $\mathcal{R}^{(4)}$ by measuring CAR$^{(4)}$ for the anti-PT system as the pump relative phase $\theta$ is tuned via the heater, plotted alongside the Hermitian coherent model for comparison. The ratio is highly tunable for the anti-PT case, whereas the coherent system shows nearly independent photon pairs ($\mathcal{R}^{(4)}\approx1$), with regimes of anti-correlation ($\mathcal{R}^{(4)}<1$) that suppress four-photon events. The calibration details mapping the heater power to the tuning phase are provided in Supplementary Information Section S4. Furthermore, we compare the measured four-photon coincidence counts (after subtracting the mean accidental background) with the theoretical four-photon correlation $G^{(4)}=\langle a_s^{\dagger}a_i^{\dagger}b_s^{\dagger}b_i^{\dagger}b_ib_sa_ia_s\rangle$ as a function of $\theta$, as shown in Fig.~\ref{four_data}(c). The four-photon correlation reaches its maximum at $\theta=0$ and its minimum at $\theta=\pi$. The visibility, defined as $\mathcal{V}=(G^{(4)}_{\rm max}-G^{(4)}_{\rm min})/(G^{(4)}_{\rm max}+G^{(4)}_{\rm min})$, is 
$94\%$ in the experiment, compared with $99\%$ predicted by the theoretical model. 

In addition, we measure three-photon coincidences $C_{s_{\mu}, s_{\nu};s_{\lambda}}$, where $s_{\mu}, s_{\nu};s_{\lambda}\in\{(a_s, a_i; b_s), (a_s, a_i;b_i), \\(b_s, b_i;a_s), (b_s, b_i; a_i)\}$. These correspond to coincidences between a virtual channel defined by a signal-idler pair $(s_{\mu},s_{\nu})$ in one waveguide and a single mode $s_{\lambda}$ in the other waveguide. An example of the three-photon coincidences, $C_{a_s, a_i;b_i}$, is shown in Fig.~\ref{three_data}(a). Similarly to $\mathcal{R}^{(4)}$, to characterize the nontrivial three-photon correlations among the four modes, we define the generalized three-photon inter-pair correlation ratio as 
$\mathcal{R}^{(3)}_{s_{\mu},s_{\nu};s_{\lambda}} = \langle s^{\dagger}_{\mu}s^{\dagger}_{\nu}s^{\dagger}_{\lambda}s_{\lambda} s_{\nu} s_{\mu}\rangle/(\langle s_{\mu}^{\dagger}s_{\nu}^{\dagger}s_{\nu} s_{\mu}\rangle \langle s_{\lambda}^{\dagger}s_{\lambda}\rangle)$. Here, the denominator normalizes the three-fold correlation by the product of the two-photon correlation of the signal-idler pair $(s_\mu,s_\nu)$ and the mean photon number of mode $s_\lambda$. The phase tuning of $\mathcal{R}^{(3)}$ of the three-photon coincidences, $C_{a_s,a_i;b_i}$, is shown in Fig.~\ref{three_data}(b).  The unnormalized three-photon correlation $G^{(3)}=\langle a_s^{\dagger}a_i^{\dagger}b^{\dagger}_ib_ia_ia_s\rangle$ as a function of $\theta$ is provided in Fig.~\ref{three_data}(c). The corresponding theoretical and experimental visibilities are $98\%$ and $92\%$, respectively.

At the two-photon level, we measure coincidences $C_{s_j,s_k}$, where $s_j,s_k\in\{(a_s,a_i),(b_s,b_i),(a_i,b_s),(a_s,b_i)\}$. Representative histograms of two-photon coincidences are presented in Fig.~\ref{two_data}(a). The differences in peak values are attributed to unequal propagation losses among the channels. The phase tuning of CAR$^{(2)}$ values, characterizing the purity of the two-photon coincidences, $C_{s_j,s_k}$, is shown in Fig.~\ref{two_data}(b). As can be seen, the purities of inter-waveguide photon pair correlations, CAR$_{a_i,b_s}$ and CAR$_{a_s,b_i}$, exhibit obvious phase-dependent tuning, whereas the purities of intra-waveguide pair correlations, CAR$_{a_s,a_i}$ and CAR$_{b_s,b_i}$, remain nearly unaffected by the phase. The origin of these nontrivial correlations can be understood from the selective action of loss on the bright and dark modes, as revealed by Eqs.~(\ref{NL}) and (\ref{BDmaster}). At $\theta=0$, the nonlinear Hamiltonian yields $H_\mathrm{NL}(0) \propto B_s^\dagger B_i^\dagger + D_s^\dagger D_i^\dagger$, allowing for the generation of protected dark-mode pairs $D_{s}^{\dagger}D_i^{\dagger} = \frac{1}{2}(a_s^\dagger a_i^\dagger - a_s^\dagger b_i^\dagger - b_s^\dagger a_i^\dagger + b_s^\dagger b_i^\dagger)$, while the bright-mode pairs $B_{s}^{\dagger}B_i^{\dagger} = \frac{1}{2}(a_s^\dagger a_i^\dagger + a_s^\dagger b_i^\dagger + b_s^\dagger a_i^\dagger + b_s^\dagger b_i^\dagger)$ are suppressed during the dissipative evolution. As a result, the inter-waveguide correlations associated with $a_s^{\dagger}b_i^{\dagger}$ and $b_s^{\dagger}a_i^{\dagger}$ can survive. However, at $\theta=\pi$, the generation shifts entirely to the cross-correlated bright/dark pairs: $H_\mathrm{NL}(\pi) \propto B_s^\dagger D_i^\dagger + D_s^\dagger B_i^\dagger = a_s^\dagger a_i^\dagger - b_s^\dagger b_i^\dagger$. Because each of these pairs contains one dissipative bright-mode photon, the contributions to the inter-waveguide correlations cancel. Therefore, the intra-waveguide terms $a_s^\dagger a_i^\dagger$ and $b_s^\dagger b_i^\dagger$ remain insensitive to the phase, whereas the survival of the inter-waveguide correlations depends strongly on the phase. This accounts for the phase-sensitive purity of the inter-waveguide pair correlations, as well as the observed phase tuning of the nontrivial inter-pair ratios $\mathcal{R}^{(3)}$ and $\mathcal{R}^{(4)}$. 

\section{Discussion}\label{sec4}
We have demonstrated a nanophotonic platform for decoherence-induced generation of higher-order quantum correlations. By linking two SPDC processes through their coupling to a common lossy reservoir, we have created exotic quantum states among two, three, and four photons. In this paradigm, decoherence, rather than causing loss of quantum information, serves as a resource to robustly create quantum correlations. This robustness lends our approach to generation of quantum correlated and entangled states over even higher photon numbers. Our results highlight a little-explored territory in quantum optics and information processing, where loss and decoherence are employed to accomplish challenging tasks. This counterintuitive realization invites further studies into re-examining the notion of loss and decoherence in quantum systems, potentially leading to striking quantum technology and devices.

\section{Methods}\label{sec11}
\textbf{Device Fabrication}

\noindent \textbf{Waveguide} The devices were fabricated from a commercial 600-nm z-cut TFLN with a 2-$\mu$m SiO$_2$ and 0.5-mm Si substrate. Waveguide patterns were defined using electron-beam lithography (EBL, Elionix ELS-G100, 100 keV) with a hydrogen silsesquioxane (HSQ) resist mask exposed at a current of 2 nA. Following development in tetramethylammonium hydroxide (TMAH) for 30 s, the pattern was transferred into the LN layer via inductively coupled plasma etching (ICP, Oxford PlasmaPro System 100 Cobra) to an etch depth of 350 nm. Post-etch cleaning was performed using an RCA-I solution to remove LN redeposition, followed by a buffered oxide etch (BOE, 6:1) to strip the remaining HSQ mask. A thin cladding layer of 100-nm SiO$_2$ was subsequently deposited by plasma-enhanced chemical vapor deposition (PECVD, PlasmaPro NGP80).

\noindent\textbf{Periodic Poling} To fabricate electrodes for periodic poling, a bilayer resist stack consisting of PMMA 495A6 and PMMA 950A4 (total thickness of 660 nm) was spin-coated onto the chip. The electrode patterns were defined using a 50 keV EBL system (Elionix ELS-LS50) and developed in a 1:1 solution of IPA:MIBK. A metal stack of 30/60-nm Cr/Au was deposited using an electron-beam evaporator (Angstrom Ultra High Vacuum Nexdep), followed by lift-off in Remover PG. Ferroelectric domain inversion was achieved by applying 60 cycles of 480-V high-voltage pulses to the electrodes. The metal electrodes were subsequently removed using Cr/Au etchants.

\noindent\textbf{Cr Strip and Heaters} Following the poling process, the Cr metal strip (250-nm wide, 20-nm thick) was fabricated on the 100-nm SiO$_2$ cladding layer, aligned directly above the central waveguide, using the same lithography and deposition procedures as the periodic poling electrodes. The device was then encased in a 1.4-$\mu$m thick SiO$_2$ upper cladding layer via PECVD. Finally, the heaters were patterned using a bilayer mask of PMMA 495A11 and PMMA 950A4 (total thickness of 1.5 $\mu$m) and 50 keV EBL. After development in 1:1 IPA:MIBK, a 15/85-nm Ti/Pt layer was deposited and patterned via lift-off in Remover PG.\\

\noindent\textbf{Evaluation of the Nonlinear Coupling Constant}

\noindent Here we provide the details of the analytical evaluation of the parametric interaction strength $g\epsilon$ used in the numerical simulations. The normalized nonlinear conversion coefficient for parametric processes is defined as \cite{g}
\begin{equation}
    g=\sqrt{\frac{16\pi^3}{\epsilon_0n_{\omega}^4n_{2\omega}^2\lambda_{\omega}^2\lambda_{2\omega}}}\frac{d_{\rm eff}\zeta}{\sqrt{A_{\rm eff}}},
\end{equation}
where $\epsilon_0$ is the vacuum permittivity, $n_{\omega}$ and $n_{2\omega}$ are the refractive indices at the fundamental and second-harmonic frequencies, and $\lambda_{\omega}$ and $\lambda_{2\omega}$ are the corresponding wavelengths. The spatial mode overlap factor is 
\begin{equation}
    \zeta=\frac{\int (E_{\omega}^{*})^2E_{2\omega}dydz}{\biggl|\int |\textbf{E}_{\omega}|^2\textbf{E}_{\omega}dydz\biggr|^{\frac{2}{3}}\biggl|\int |\textbf{E}_{2\omega}|^2\textbf{E}_{2\omega}dydz\biggr|^{\frac{1}{3}}}.
\end{equation}
The effective mode area is defined as $A_{\rm eff}=(A_{\omega}^2A_{2\omega})^{\frac{1}{3}}$, where $A_\mu=(\int |\textbf{E}_{\mu}|^2dydz)^3/\bigl|\int |\textbf{E}_\mu|^2\textbf{E}_\mu dydz\bigr|^2$, $\mu=\omega,2\omega$. The transverse electric field profiles of the fundamental and second-harmonic TM$_{00}$
 modes were obtained from the FDE solver in Lumerical MODE and exported to MATLAB for numerical evaluation of the above quantities, yielding $\zeta=0.92$ and $A_{\rm eff}=1.11$ $\mu m^2$. Using $\lambda_{\omega}=1550$ nm, $\lambda_{2\omega}=775$ nm, $n_{\omega}=1.8926$, $n_{2\omega}=2.1265$, and the effective nonlinear
coefficient $d_{\rm eff}=\frac{2}{\pi}d_{33}=17.19$ pm/V, where $d_{33}$ is the largest second-order nonlinear tensor element of the z-cut TFLN, one obtains $g=1.08\times10^{10}$ m$^{-1}$J$^{-1/2}$. The analytical value is compared with the nonlinear conversion coefficient extracted from the SHG characterization experiment on a reference waveguide (Fig.~\ref{fig1}(c)), which shares the same geometry and poling-period design as the triple anti-PT waveguides. The coefficient is defined by
\begin{equation}
    g_{\text{exp}}=\sqrt{\frac{P_{2\omega}(L)}{P_{\omega}(0)}\frac{2\pi c\lambda_{2\omega}}{L^2\lambda_{\omega}^2}}.
\end{equation}
The fundamental power at the input of the chip was $P_{\omega}(0)=2.8\times 10^{-3}$ W, and the generated second-harmonic power at the end of the periodically poled waveguide of length $L=4$ mm was $P_{2\omega}(L)=1.74\times 10^{-5}$ W. The experimentally determined nonlinear conversion coefficient is therefore $g_{\text{exp}}=9.18\times10^{9}$ m$^{-1}$J$^{-1/2}$, in close agreement with the analytical value.

The pump in the SPDC experiment is treated as a classical undepleted field with amplitude $\epsilon$ related to the on-chip pump power by $\epsilon=\sqrt{\lambda_{2\omega}P_{2\omega}(0)/(8\pi c)}$. For an on-chip pump power $P_{2\omega}(0)=4\times 10^{-3}$ W, this yields $\epsilon=6.41\times 10^{-10}$ J$^{1/2}$. Hence, the corresponding pair-generation coupling strength per unit length is $g\epsilon=6.92$ m$^{-1}$.\\

\noindent\textbf{Experiment Details}

\noindent In the SHG characterization experiment (Fig.~\ref{setup}), the fundamental wavelength laser (Santec) is connected to the fundamental input port of a Fused Wavelength Division Multiplexer (FWDM, OF-LINK). Next, we measure the SHG power at each output fiber using a power meter (Thorlabs PM100D) while sweeping the input fundamental wavelength from 1500 to 1630 nm. The anti-PT chip is mounted on a stage equipped with a thermoelectric cooler (TEC, Vescent Photonics, Slice-QTC) that actively stabilizes the temperature. We determined the optimal operating temperature to be 69 \textdegree C, which maximizes the SHG output power at the QPM wavelength (reaching 30 nW after Cr absorption).

In the central SPDC experiment, a CW 775.4-nm laser (New Focus TLB-6700) is launched into a fiber (Thorlabs P5-630A-PCAPC-1) and the polarization is tuned by a fiber polarization controller (FPC, Thorlabs). The light is then fed into the 780-nm input port of an FWDM. The FWDM output is coupled to a lensed fiber (OZ Optics) that focuses the beam to a spot size of 2 $\mu$m to match the waveguide mode size, resulting in a fiber-to-chip coupling loss of 5 dB per facet. The sample temperature is stabilized by the TEC at 69 \textdegree C with a stability of $\pm 3$ mK. The separation between the two output waveguide paths is designed to be 127 $\mu$m; this matches the 127-$\mu$m pitch of the output lensed fiber array (OZ Optics). Each output path incorporates two cascaded long-pass filters (Thorlabs FELH0950), providing a total of 140-dB attenuation of the pump light. Along each path, two cascaded narrow-band Dense Wavelength Division Multiplexing (DWDM) filters (AC Photonics, 125-GHz bandwidth) centered at 1549.35 nm and 1552.54 nm are employed to separate the signal and idler photons, respectively. The four output fibers are connected to the SNSPDs (ID Quantique ID281, detection efficiency of $85\%$, dark count of 50-100 Hz). The converted electric pulses are sent to the Time Tagger (Swabian Instruments) to perform coincidence counting. 

To compare with the theoretical prediction of $g^{(4)}$, we define an experimentally useful expression for normalized four-photon correlation as \cite{wolf}
\begin{equation}
    g^{(4)}_{\text{exp}}=\frac{C_4(\tau_1,\tau_2,\tau_3)}{R_1R_2R_3R_4 T\Delta\tau_1\Delta\tau_2\Delta\tau_3},
\end{equation}
where $C_4(\tau_1,\tau_2,\tau_3)$ represents the four-photon coincidence counts as a function of the relative time delays: $\tau_1$ between channels $a_s$ and $a_i$, $\tau_2$ between channels $b_s$ and $b_i$, and $\tau_3$ between the two virtual channels. $R_{i}$ are the single count rates for each channel, $T=30$ min is the total integration time, and $\Delta\tau_{j}$ represent the coincidence-window widths for the three independent temporal coordinates. For experimental measurements of $g^{(4)}$ versus $\theta$, see Supplementary Information Section S3.



\backmatter

\bmhead{Acknowledgements}
YPH sincerely thanks Jianming Wen for enlightening discussions and enjoyable collaboration that motivated this research effort.

\bmhead{Funding}
This research was supported in part by ACC-New Jersey under grant number W15QKN18D0040 and Office of Naval Research under grant number N00014-21-1-2898. Device fabrication was performed at Advanced Science Research Center (ASRC) at the City University of New York (CUNY) and the Columbia Nano Initiative (CNI) at Columbia University.

\bmhead{Data availability}
The data underlying this study are available from the corresponding authors upon request.

\bmhead{Competing interests}
The authors declare no competing interests.





\bibliography{sn-bibliography}

\end{document}


\maketitle

\section{Supplementary Section S1.~Chip Design}
For the second harmonic generation (SHG) or spontaneous parametric down-conversion (SPDC) process, the poling period $\Lambda$ can be determined by the quasi-phase-matching (QPM) condition
\begin{equation}
    \Delta k=\frac{2\pi n_{775}}{\lambda_{775}}-\frac{4\pi n_{1550}}{\lambda_{1550}}-\frac{2\pi}{\Lambda}=0,
\end{equation}
where the subscripts denote wavelengths in nanometers. We performed eigenmode simulations (FDE Solver, Lumerical MODE) using the waveguide geometry shown in Fig.~1(b) of the main text. The solution yields effective refractive indices of $n_{1550}=1.8875$ and $n_{775}=2.1262$ for the fundamental and second-harmonic TM$_{00}$ modes, respectively. The corresponding QPM poling period is therefore $\Lambda=3.25$ $\mu$m.

\begin{figure}[h]
\centering
\hspace*{0.01\linewidth}
\begin{tikzpicture}
\node[anchor=south west, inner sep=0] (img) at (0,0)
    {\includegraphics[width=0.9\linewidth]{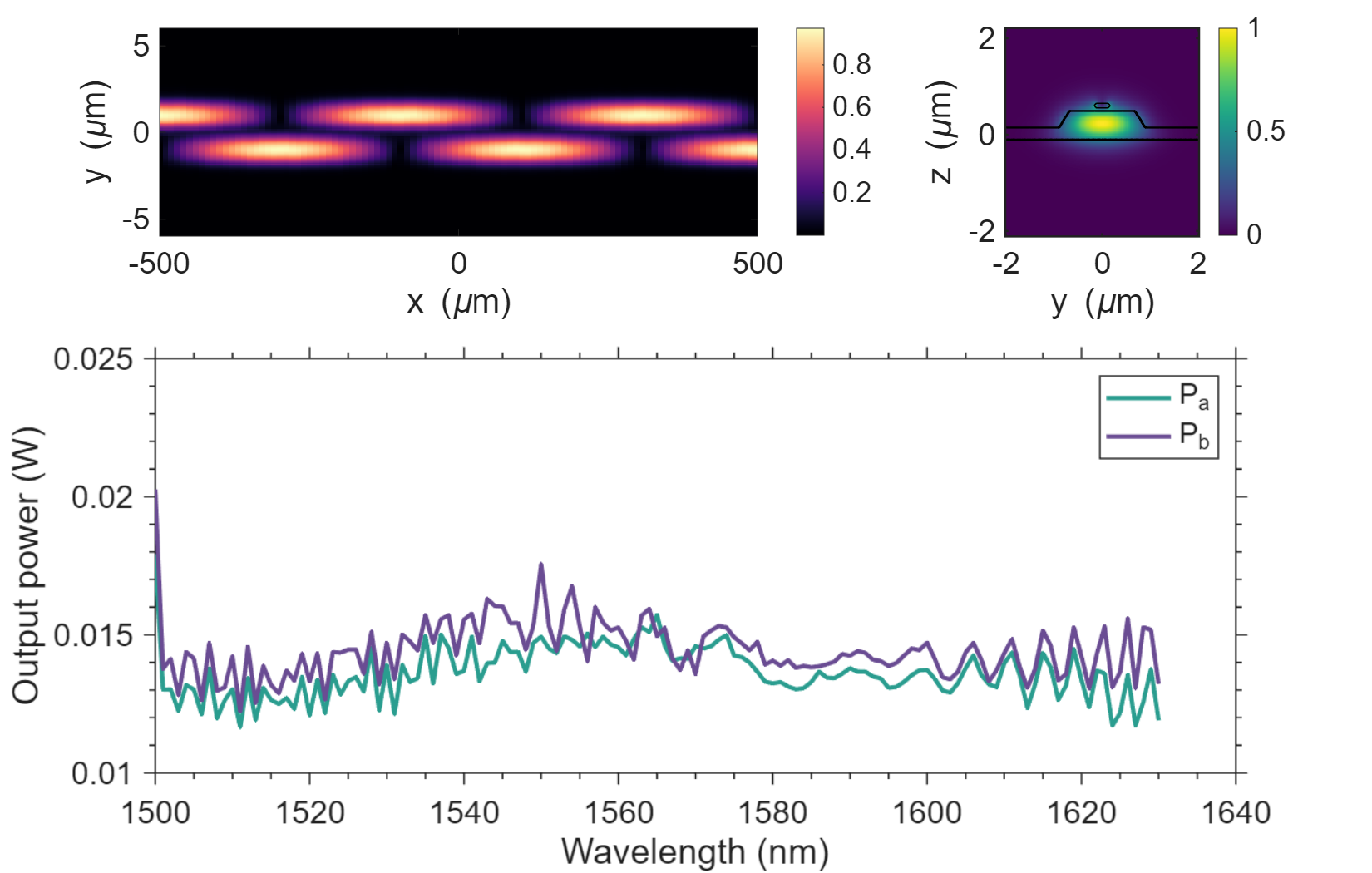}};

  \path[use as bounding box] (img.south west) rectangle (img.north east);
\node at (1.1,9.5) {\textbf{(a)}};
\node at (10.2,9.5) {\textbf{(b)}};
\node at (1.1,6.1) {\textbf{(c)}};
\end{tikzpicture}

\caption{\textbf{Engineering of the effective coupling strength $\Gamma$ via the design of the hopping rate $\kappa$ and the Cr-induced dissipation rate $\gamma$}. (a) EME simulation of light propagation in coupled waveguides; the color scale represents the electric-field amplitude.
(b) Simulated transverse electric-field amplitude profile illustrating distortion caused by absorption in the Cr metal strip.
(c) Experimental verification of equal power splitting, a signature of operation in the anti-PT symmetric phase.}\label{fig1}
\end{figure}

The hopping rate $\kappa$ between adjacent waveguides and the dissipation rate $\gamma$ of the waveguide $c$ are calculated from numerical simulations performed using EME and FDE Solvers in Lumerical MODE, respectively, from which the effective coupling $\Gamma=|\kappa|^2/\gamma$ is obtained. The gap between adjacent waveguides is set to 600 nm. The EME propagation simulation result for 1550-nm light is shown in Fig.~S1(a). The beating length is $L_{\rm beat}=205$ $\mu$m, defined as the distance corresponding to the maximum power transfer between the two waveguides. The hopping rate is therefore given by
\begin{equation}
    \kappa_{1550}=\frac{\pi}{2L_{\rm beat}}=76.62\ \rm cm^{-1}.
\end{equation}
Applying the same procedure, we obtain a hopping rate of $\kappa_{775}=3.14$ cm$^{-1}$ for 775-nm light. The weak coupling condition for 775-nm light, $\kappa_{1550}\gg\kappa_{775}$, ensures that the pump coupled into waveguide $c$ is negligible ($\epsilon_c\approx 0$), as mentioned in the main text. 

The dissipation rate $\gamma$ induced by the Cr strip is simulated using the FDE Solver in Lumerical MODE. Using the geometry parameters shown in Fig.~1(b) of the main text, we find $\gamma_{1550}=813$ cm$^{-1}$ and $\gamma_{775}=14$ cm$^{-1}$. The transverse field profile is provided in Fig.~S1(b).

The effective coupling strength for 1550-nm light is $\Gamma=|\kappa|^2/\gamma=7.22$ cm$^{-1}$. Since the two waveguides $a$ and $b$ are designed to be identical, we assume the effective detuning $\Delta=(k_a-k_b)/2\approx0$. With the condition $\Gamma>\Delta$, the four-photon correlation is investigated in the anti-PT symmetric phase. As a characteristic feature of anti-PT symmetry, equal power splitting\cite{Anti_PT1,Anti_PT2}between the outputs of waveguides $a$ and $b$ for classical light over the spectral range $1500 - 1630$ nm is verified experimentally (Fig.~S1(c)), confirming that the system operates in the anti-PT symmetry regime.

\begin{figure}[h]
\centering
\hspace*{0.01\linewidth}
\begin{tikzpicture}
\node[anchor=south west, inner sep=0] (img) at (0,0)
    {\includegraphics[width=0.9\linewidth]{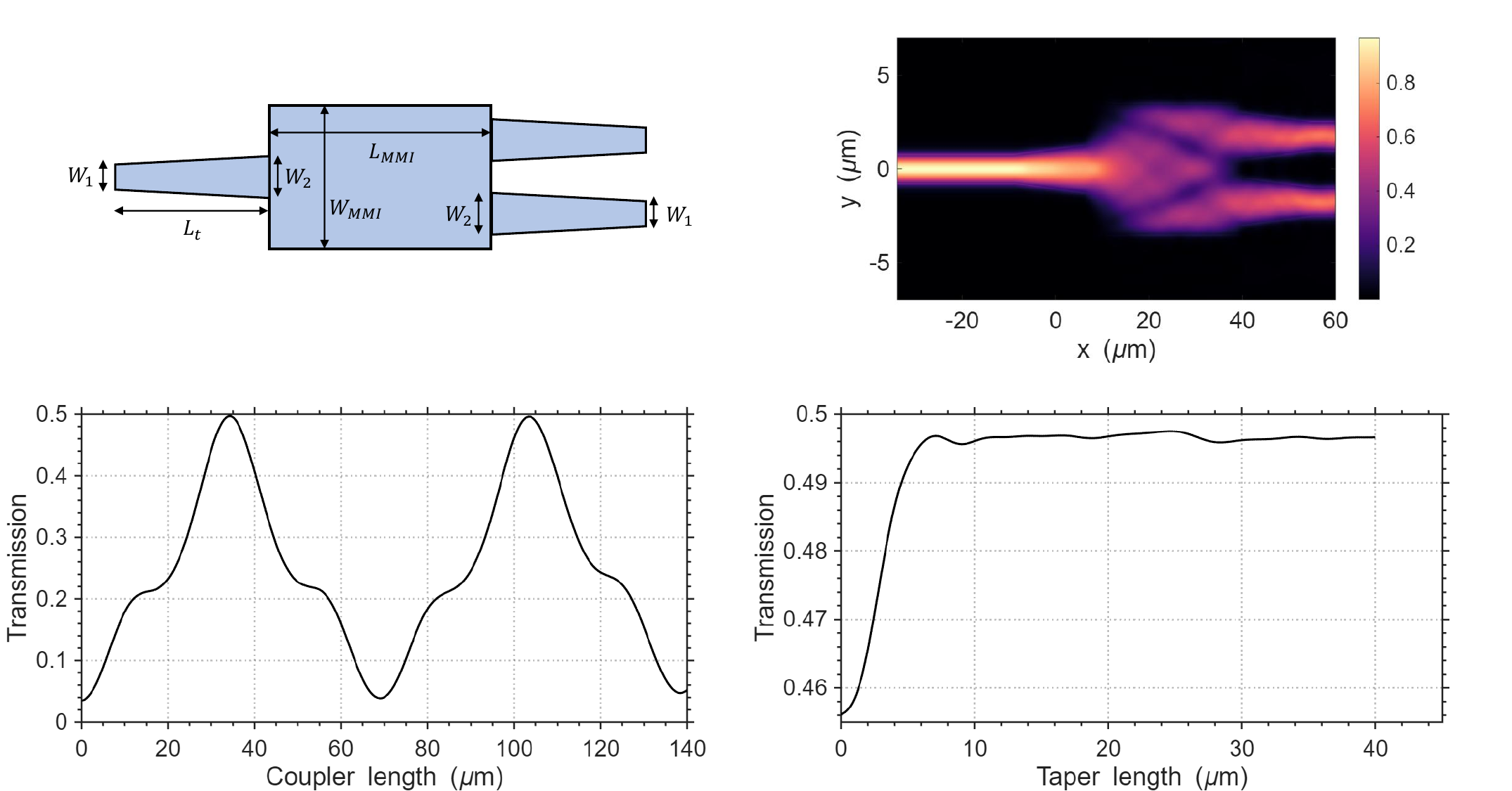}};

  \path[use as bounding box] (img.south west) rectangle (img.north east);
\node at (0.5,7.7) {\textbf{(a)}};
\node at (8.2,7.7) {\textbf{(b)}};
\node at (0.5,4.2) {\textbf{(c)}};
\node at (8.2,4.2) {\textbf{(d)}};
\end{tikzpicture}

\caption{\textbf{Multimode interferometer (MMI) design for the MZI}. (a) Schematic of the MMI coupler geometry. (b) Electric field distribution in the MMI from EME simulation. (c) Simulated transmission at one MMI output port versus the coupler length $L_{\rm MMI}$, showing an optimized value of $L_{\rm MMI}=34$ $\mu$m. (d) Simulated transmission at one MMI output port versus the taper length $L_{\rm t}$, which plateaus for $L_{\rm t}>7$ $\mu$m. Accordingly, $L_{\rm t}=10$ $\mu$m is chosen for the MMI taper design.}\label{fig1}
\end{figure}

The input splitting junction of the Mach–Zehnder interferometer (MZI) is realized using a $1\times 2$ multimode interferometer (MMI; structure shown in Fig.~S2(a)). The MMI geometry is optimized using EME Solver in Lumerical MODE. As shown in Fig.~S2(c, d), the MMI coupler length $L_{\rm MMI}$ and the taper length $L_{\rm t}$ are varied to identify the optimized geometry corresponding to maximum transmission. The optimized values are found to be $L_{\rm MMI}=34$ $\mu$m and $L_{\rm t}=10$ $\mu$m. The corresponding electric field distribution within the MMI is shown in Fig.~S2(b).

\section{Supplementary Section S2.~Analytical Solutions of Anti-PT and Coherent Systems at Phases $\theta=0$ and $\pi$}
To make the physics transparent, we use the bright/dark ($B/D$) mode basis, defined as 
\begin{equation}
    B_{\mu}=\frac{a_{\mu}+b_{\mu}}{\sqrt{2}},\ \ D_{\mu}=\frac{a_{\mu}-b_{\mu}}{\sqrt{2}},\ \ \ (\mu=s,i).
\end{equation}
In this basis, the Hamiltonian of the coherent system reads
\begin{equation}
    H_{\rm c}=H_{\rm NL}+\Gamma\sum_{\mu=s,i}(B_{\mu}^{\dagger}B_{\mu}-D_{\mu}^{\dagger}D_{\mu}),
\end{equation}
where the SPDC interaction term is $H_{\rm NL}(B,D)=\Lambda_{\text{co}}(B_s^{\dagger}B_i^{\dagger}+D_s^{\dagger}D_i^{\dagger})+\Lambda_{\text{x}}(B_s^{\dagger}D_i^{\dagger}+D_s^{\dagger}B_i^{\dagger})+\text{h.c.}$ The co-generation and cross-generation coefficients are  $\Lambda_{\text{co}}=g\epsilon e^{i\theta/2}\cos(\theta/2)$ and $\Lambda_{\text{x}}=ig\epsilon e^{i\theta/2}\sin(\theta/2)$, respectively. In the Heisenberg picture, the equations of motion for the annihilation operators are
\begin{equation}
    \frac{dB_s}{dz} = -i\Gamma B_s
-i(\Lambda_{\rm co}B_i^{\dagger}+\Lambda_{\rm x}D_i^{\dagger}),
\end{equation}
\begin{equation}
    \frac{dD_s}{dz} = i\Gamma D_s
-i(\Lambda_{\rm co}D_i^{\dagger}+\Lambda_{\rm x}B_i^{\dagger}),
\end{equation}
\begin{equation}
    \frac{dB_i}{dz} = -i\Gamma B_i
-i(\Lambda_{\rm co}B_s^{\dagger}+\Lambda_{\rm x}D_s^{\dagger}),
\end{equation}
\begin{equation}
    \frac{dD_i}{dz} = i\Gamma D_i
-i(\Lambda_{\rm co}D_s^{\dagger}+\Lambda_{\rm x}B_s^{\dagger}).
\end{equation}
And the adjoint equations follow directly from Hermitian conjugation; e.g.,
\begin{equation}
    \frac{dB_s^{\dagger}}{dz} = i\Gamma B_s^{\dagger}
+i(\Lambda^*_{\rm co}B_i+\Lambda^*_{\rm x}D_i),
\end{equation}
etc. 

When $\theta=0$, the coefficients are $\Lambda_{\rm co}=g\epsilon$ and  $\Lambda_{\rm x}=0$. Since $\Gamma>g\epsilon$, we introduce $\Omega=\sqrt{\Gamma^2-(g\epsilon)^2}$. By transforming to the waveguide mode basis, the evolution of the four-mode operator vector $\hat{\mathbf{V}}(z) = [a_s(z), b_s(z), a_i^\dagger(z), b_i^\dagger(z)]^T$ is governed by the Bogoliubov transformation:
\begin{equation}
    \hat{\mathbf{V}}(z) = \mathbf{M}_1(z) \hat{\mathbf{V}}(0),
\end{equation}
where the transfer matrix is given by
\begin{equation}
    \textbf{M}_1(z)=\begin{pmatrix}
\cos(\Omega z) & -i \frac{\Gamma}{\Omega} \sin(\Omega z) & -i \frac{g\epsilon}{\Omega}\sin(\Omega z) & 0 \\
-i \frac{\Gamma}{\Omega} \sin(\Omega z) & \cos(\Omega z) & 0 & -i \frac{g\epsilon}{\Omega} \sin(\Omega z) \\
i \frac{g\epsilon}{\Omega} \sin(\Omega z) & 0 & \cos(\Omega z) & i \frac{\Gamma}{\Omega} \sin(\Omega z) \\
0 & i \frac{g\epsilon}{\Omega}\sin(\Omega z) & i \frac{\Gamma}{\Omega} \sin(\Omega z) & \cos(\Omega z)
\end{pmatrix}.
\end{equation}
From the solutions above, the mean photon number (which is identical for all four modes) is given by
\begin{equation}
    \langle n(z)\rangle=\langle 0|\hat{n}(z)|0\rangle=\biggl|-i\frac{g\epsilon}{\Omega}\sin(\Omega z)\biggr|^2=\frac{g^2\epsilon^2}{\Omega^2}\sin^2(\Omega z).
\end{equation}
Defining $u = \cos(\Omega z)$, $v = -i \frac{\Gamma}{\Omega} \sin(\Omega z)$, and $w = -i \frac{g\epsilon}{\Omega} \sin(\Omega z)$, the unnormalized four-photon correlation function is given by
\begin{equation}
\begin{split}
        G^{(4)}(z)&=\langle a_s^{\dagger}a_i^{\dagger}b_s^{\dagger}b_i^{\dagger}b_ib_sa_ia_s\rangle=|w|^4|u^2+v^2|^2+2|w|^6(|u|^2+|v|^2)+|w|^8\\
        &=\biggl(\frac{g\epsilon}{\Omega}\biggr)^8\sin^8(\Omega z)\biggl(\frac{\Omega^4}{(g\epsilon)^4}\frac{\cos^4(\Omega z)}{\sin^4(\Omega z)}+\frac{\Gamma^4}{(g\epsilon)^4}-2\frac{\Omega^4}{(g\epsilon)^4}\frac{\cos^2(\Omega z)}{\sin^2(\Omega z)}\\
        &+2\frac{\Gamma^2}{(g\epsilon)^2}+1\biggr)=\biggl(\frac{g\epsilon}{\Omega}\biggr)^8\sin^8(\Omega z)\biggl(\frac{4\Gamma^2}{(g\epsilon)^2}+\frac{\Omega^4}{(g\epsilon)^4}\frac{\cos^2(2\Omega z)}{\sin^4(\Omega z)}\biggr).
\end{split}
\end{equation}

For the coherent system at phase $\theta=\pi$, the coefficients reduce to $\Lambda_{\rm co}=0$ and  $\Lambda_{\rm x}=-g\epsilon$. The corresponding transfer matrix of the solution is then given by
\begin{equation}
    \textbf{M}_2(z)=
   \begin{pmatrix}
C c_\Gamma & -i C s_\Gamma & i S c_\Gamma & - S s_\Gamma \\
-i C s_\Gamma & C c_\Gamma & - S s_\Gamma & i S c_\Gamma \\
-i S c_\Gamma & S s_\Gamma & C c_\Gamma & i C s_\Gamma \\
S s_\Gamma & -i S c_\Gamma & i C s_\Gamma & C c_\Gamma
\end{pmatrix},
\end{equation}
where $C = \cosh(g\epsilon z)$, $S = \sinh(g\epsilon z)$,  $c_\Gamma = \cos(\Gamma z)$, and $s_\Gamma = \sin(\Gamma z)$. The mean photon number is given by
\begin{equation}
    \langle n(z)\rangle=\sinh^2(g\epsilon z),
\end{equation}
and we obtain
\begin{equation}
    G^{(4)}(z)=S^4C^4+2S^6C^2+S^8=\sinh^4(g\epsilon z)\cosh^2(2g\epsilon z).
\end{equation}
The analytical results for $G^{(4)}(z)$ in the Hermitian coherent system at the two specific phases are compared with the corresponding numerical simulations, as shown in Fig.~S3(a,b).\\
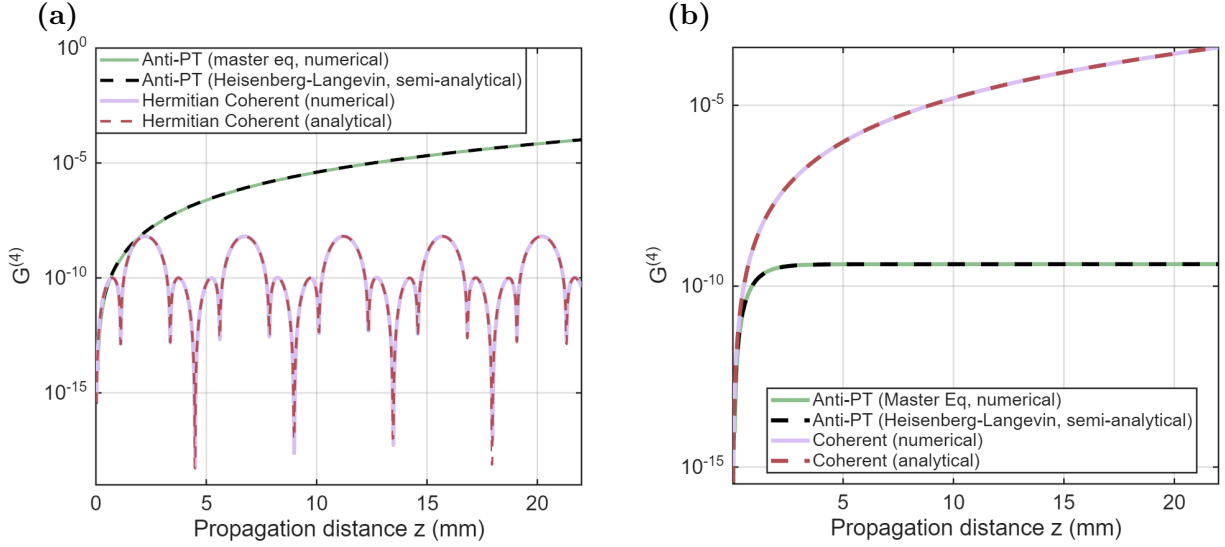
\begin{figure}[H]
\centering

\begin{tikzpicture}
\node[anchor=south west, inner sep=0] (img) at (0,0)
    {\includegraphics[width=\linewidth]{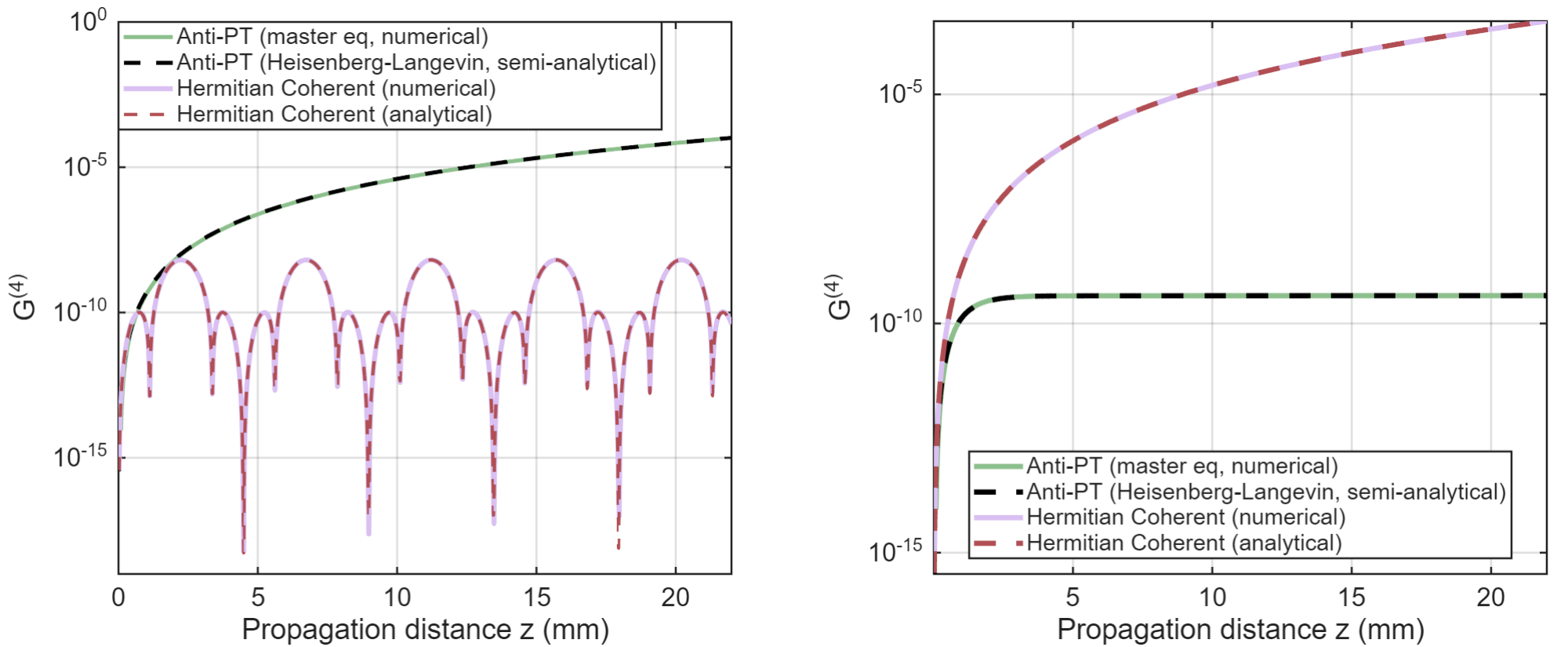}};

  \path[use as bounding box] (img.south west) rectangle (img.north east);
\node at (0.8,7) {\textbf{(a)}};
\node at (9.3,7) {\textbf{(b)}};
\end{tikzpicture}

\caption{\textbf{Comparison of numerical and analytical} $G^{(4)}(z)$ \textbf{dynamics in anti-PT and Hermitian coherent systems}. Results are shown for two specific input relative phases: (a) $\theta=0$ and (b) $\theta=\pi$. Numerical and analytical results agree well in both cases.} \label{figS3}
\end{figure}

For the anti-PT system, the Heisenberg-Langevin equations of motion are 
\begin{equation}\label{HL1}
    \frac{dB_s}{dz} = -2\Gamma\,B_s-i\Big(\Lambda_{\rm co}B_i^\dagger+\Lambda_{\rm x}D_i^\dagger\Big)+\sqrt{4\Gamma}\,B_{s,\mathrm{in}}(z),
\end{equation}
\begin{equation}\label{HL2}
    \frac{dD_s}{dz}= - i\Big(\Lambda_{\rm co}D_i^\dagger+\Lambda_{\rm x}B_i^\dagger\Big),
\end{equation}
\begin{equation}\label{HL3}
    \frac{dB_i}{dz} = -2\Gamma\,B_i- i\Big(\Lambda_{\rm co}B_s^\dagger+\Lambda_{\rm x}D_s^\dagger\Big)
+\sqrt{4\Gamma}B_{i,\mathrm{in}}(z),
\end{equation}
\begin{equation}\label{HL4}
    \frac{dD_i}{dz} = - i\Big(\Lambda_{\rm co}D_s^\dagger+\Lambda_{\rm x}B_s^\dagger\Big),
\end{equation}
together with the Hermitian conjugate of these equations. To preserve the commutation relation $[B(z),B^{\dagger}(z)]=1$ during propagation, vacuum input noise operators $B_{\rm in}(z)$ are included, with $[B_{\rm in}(z),B_{\rm in}^{\dagger}(z)]=1$ \cite{QN1,QN2}.

By defining $\hat{\textbf{W}}=(B_s,D_s,B_i^{\dagger},D_i^{\dagger})^T$and $\hat{\textbf{N}}=\sqrt{4\Gamma}(B_{s,\rm in},0,B^{\dagger}_{i,\rm in},0)^{T}$, Eqs.~(\ref{HL1})-(\ref{HL4}) can be written compactly in the bright/dark basis as
\begin{equation}
    \frac{d\hat{\textbf{W}}(z)}{dz}=\textbf{A}_{\theta}\hat{\textbf{W}}(z)+\hat{\textbf{N}}(z)
\end{equation}
where 
\begin{equation}
    \textbf{A}_{\theta}=
    \begin{pmatrix}
    -2\Gamma & 0 & -i\Lambda_{\rm co} & -i\Lambda_{\rm x}\\
    0 & 0 & -i\Lambda_{\rm x} & -i\Lambda_{\rm co}\\
    i\Lambda^*_{\rm co}&i\Lambda^*_{\rm x}&-2\Gamma& 0\\
    i\Lambda^*_{\rm x}&i\Lambda^*_{\rm co}&0&0
    \end{pmatrix}.
\end{equation}
The formal solution for an arbitrary relative pump phase $\theta$ is therefore
\begin{equation}
    \hat{\textbf{W}}(z)=\textbf{M}_{\theta}(z)\hat{\textbf{W}}(0)+\hat{\textbf{F}}_{\theta}(z),\ \ \ \ \textbf{M}_{\theta}(z)=e^{\textbf{A}_{\theta}z}
\end{equation}
with 
\begin{equation}
    \hat{\textbf{F}}_{\theta}(z)=\int_0^{z}d\zeta \textbf{M}_{\theta}(z-\zeta)\hat{\textbf{N}}(\zeta).
\end{equation}
With the initial joint system–reservoir state 
\begin{equation}
    |\Psi\rangle=|0,0;0,0\rangle_{\rm sys}\otimes|\rm vac\rangle_{\rm res},
\end{equation}
being a Gaussian vacuum state, the linear Bogoliubov–Langevin evolution preserves Gaussianity \cite{QN1, Gaussian}, and the system state (after tracing over the reservoir degrees of freedom) therefore remains Gaussian throughout the propagation.

To evaluate the four-photon correlation \(G^{(4)}\), which is defined in the physical waveguide modes \(a\) and \(b\), we transform the solution back to the waveguide basis. For \(\theta=0\), this gives 
\begin{equation}\label{solution}
    \hat{\mathbf{V}}(z) = \mathbf{M}_3(z) \hat{\mathbf{V}}(0)+\hat{\mathbf{F}}(z),
\end{equation}
where the transfer matrix is
\begin{equation}
\mathbf{M}_3(z) = \frac{1}{2}
\begin{pmatrix}
C(1 + e^{-2\Gamma z}) & -C(1 - e^{-2\Gamma z}) & -iS(1 + e^{-2\Gamma z}) & iS(1 - e^{-2\Gamma z}) \\
-C(1 - e^{-2\Gamma z}) & C(1 + e^{-2\Gamma z}) & iS(1 - e^{-2\Gamma z}) & -iS(1 + e^{-2\Gamma z}) \\
iS(1 + e^{-2\Gamma z}) & -iS(1 - e^{-2\Gamma z}) & C(1 + e^{-2\Gamma z}) & -C(1 - e^{-2\Gamma z}) \\
-iS(1 - e^{-2\Gamma z}) & iS(1 + e^{-2\Gamma z}) & -C(1 - e^{-2\Gamma z}) & C(1 + e^{-2\Gamma z})
\end{pmatrix}
\end{equation}
and the vacuum noise vector is
\begin{equation}
    \hat{\mathbf{F}}(z) = 
\begin{pmatrix}
\mathcal{F}_s \ \ 
\mathcal{F}_s \ \
\mathcal{F}_i^\dagger \ \
\mathcal{F}_i^\dagger
\end{pmatrix}^T
\end{equation}
with 
\begin{equation}
\begin{split}
        \mathcal{F}_s(z)& = \sqrt{\Gamma} \int_0^z d\zeta \, e^{-2\Gamma(z-\zeta)} \bigl[ \cosh(g\epsilon(z-\zeta)) (a_{s,\text{in}}(\zeta) + b_{s,\text{in}}(\zeta))\\
    &- i \sinh(g\epsilon(z-\zeta)) (a_{i,\text{in}}^\dagger(\zeta) + b_{i,\text{in}}^\dagger(\zeta)) \bigr]
\end{split}
\end{equation}
and
\begin{equation}
\begin{split}
     \mathcal{F}_i^\dagger(z) &= \sqrt{\Gamma} \int_0^z d\zeta \, e^{-2\Gamma(z-\zeta)} \bigl[ \cosh(g\epsilon(z-\zeta)) (a_{i,\text{in}}^\dagger(\zeta)+b_{i,\text{in}}^\dagger(\zeta)) \\
     & + i \sinh(g\epsilon(z-\zeta)) (a_{s,\text{in}}(\zeta) + b_{s,\text{in}}(\zeta))\bigr].
\end{split}
\end{equation}
The mean photon number is then given by
\begin{equation}
    \langle n(z)\rangle=\frac{1}{2}\sinh^2(g\epsilon z)\bigl(1+e^{-4\Gamma z}\bigr)+\langle n\rangle_{\rm vac},
\end{equation}
where the vacuum noise contribution is
\begin{equation}
\begin{split}
       \langle n\rangle_{\rm vac}&=\langle\mathcal{F}^{\dagger}_s\mathcal{F}_s\rangle=4\Gamma \int_0^z d\tau \, e^{-4\Gamma(z-\tau)} \sinh^2(g\epsilon(z-\tau))\\
    &=\Gamma \left[ \frac{1 - e^{-2(2\Gamma - g\epsilon)z}}{2\Gamma - g\epsilon} + \frac{1 - e^{-2(2\Gamma + g\epsilon)z}}{2\Gamma + g\epsilon} - \frac{1 - e^{-4\Gamma z}}{\Gamma} \right].
\end{split}
\end{equation}
Due to the algebraic complexity of expanding the full four-photon correlation into a single explicit formula, we adopt a semi-analytical approach. Specifically, we compute $G^{(4)}$ numerically using the exact analytical operator solutions (Eq.~(\ref{solution})). Figure~S3(a) demonstrates excellent agreement between the semi-analytical results and full numerical simulations.

At the relative phase $\theta=\pi$, the operator solution for the anti-PT system can be written as $ \hat{\mathbf{V}}(z) = \mathbf{M}_4(z) \hat{\mathbf{V}}(0)+\hat{\mathbf{F}}(z)$, where the transfer matrix is
\begin{equation}
    \mathbf{M}_4(z)=
   e^{-\Gamma z}
\begin{pmatrix}
C' & -\dfrac{\Gamma}{k}S' & i\dfrac{g\epsilon}{k}S' & 0\\[8pt]
-\dfrac{\Gamma}{k}S' & C' & 0 & -\,i\dfrac{g\epsilon}{k}S'\\[8pt]
-\,i\dfrac{g\epsilon}{k}S' & 0 & C' & -\dfrac{\Gamma}{k}S'\\[8pt]
0 & i\dfrac{g\epsilon}{k}S' & -\dfrac{\Gamma}{k}S' & C'
\end{pmatrix}.
\end{equation}
Here, $C'=\cosh(kz)$, $S'=\sinh(kz)$, and $k=\sqrt{\Gamma^2+(g\epsilon)^2}$.
Defining the kernels 
\begin{align}
K_0(\xi)
&=
e^{-\Gamma \xi}
\left[
\cosh(k\xi)
-
\frac{\Gamma}{k}\sinh(k\xi)
\right],
\\
K_1(\xi)
&=
e^{-\Gamma \xi}
\frac{g\epsilon}{k}\sinh(k\xi),
\end{align}
the vacuum noise vector is given by 
\begin{equation}
    \hat{\mathbf{F}}(z) = 
\begin{pmatrix}
\mathcal{F}_{a_s} \ \ 
\mathcal{F}_{b_s} \  \
\mathcal{F}_{a_i}^\dagger \ \ 
\mathcal{F}_{b_i}^\dagger
\end{pmatrix}^{T}
\end{equation}
with
\begin{equation}
    \begin{split}
        \mathcal F_{a_s}(z)
&=\sqrt{\Gamma}\int_0^{z}\! d\zeta\Big[
K_0(z-\zeta)\big(\hat a_{s,\mathrm{in}}(\zeta)+\hat b_{s,\mathrm{in}}(\zeta)\big)
+i\,K_1(z-\zeta)\big(\hat a_{i,\mathrm{in}}^{\dagger}(\zeta)+\hat b_{i,\mathrm{in}}^{\dagger}(\zeta)\big)
\Big],\\[6pt]
\mathcal F_{b_s}(z)
&=\sqrt{\Gamma}\int_0^{z}\! d\zeta\Big[
K_0(z-\zeta)\big(\hat a_{s,\mathrm{in}}(\zeta)+\hat b_{s,\mathrm{in}}(\zeta)\big)
-i\,K_1(z-\zeta)\big(\hat a_{i,\mathrm{in}}^{\dagger}(\zeta)+\hat b_{i,\mathrm{in}}^{\dagger}(\zeta)\big)
\Big],\\[6pt]
\mathcal{F}_{a_i^\dagger}(z)
&=\sqrt{\Gamma}\int_0^{z}\! d\zeta\Big[
K_0(z-\zeta)\big(\hat a_{i,\mathrm{in}}^{\dagger}(\zeta)+\hat b_{i,\mathrm{in}}^{\dagger}(\zeta)\big)
-i\,K_1(z-\zeta)\big(\hat a_{s,\mathrm{in}}(\zeta)+\hat b_{s,\mathrm{in}}(\zeta)\big)
\Big],\\[6pt]
\mathcal F_{b_i^\dagger}(z)
&=\sqrt{\Gamma}\int_0^{z}\! d\zeta\Big[
K_0(z-\zeta)\big(\hat a_{i,\mathrm{in}}^{\dagger}(\zeta)+\hat b_{i,\mathrm{in}}^{\dagger}(\zeta)\big)
+i\,K_1(z-\zeta)\big(\hat a_{s,\mathrm{in}}(\zeta)+\hat b_{s,\mathrm{in}}(\zeta)\big)
\Big].
    \end{split}
\end{equation}
The mean photon number is
\begin{equation}
    \langle n(z)\rangle=\frac{(g\epsilon)^2}{k^2}\sinh^2(kz)e^{-2\Gamma z}+\langle n\rangle_{\rm vac},
\end{equation}
where the vacuum contribution is
\begin{equation}
\begin{split}
       \langle n\rangle_{\rm vac}&=2\Gamma\left(\frac{g\epsilon}{k}\right)^2
\int_{0}^{z}\! d\xi e^{-2\Gamma \xi}\sinh^2(k\xi)\\
&=-\frac{1}{2}
+\frac{e^{-2\Gamma z}}{2}
\left[
\left(\frac{\Gamma}{k}\right)^2\cosh(2kz)
+\left(\frac{\Gamma}{k}\right)\sinh(2kz)
+\left(\frac{g\epsilon}{k}\right)^2
\right]. 
\end{split}
\end{equation}
Again, the four-photon correlation evolution $G^{(4)}(z)$ is evaluated semi-analytically, showing agreement with the full numerical results in Fig.~S3(b).\\

\begin{figure}[h]
\centering

\begin{tikzpicture}
\node[anchor=south west, inner sep=0] (img) at (0,0)
    {\includegraphics[width=\linewidth]{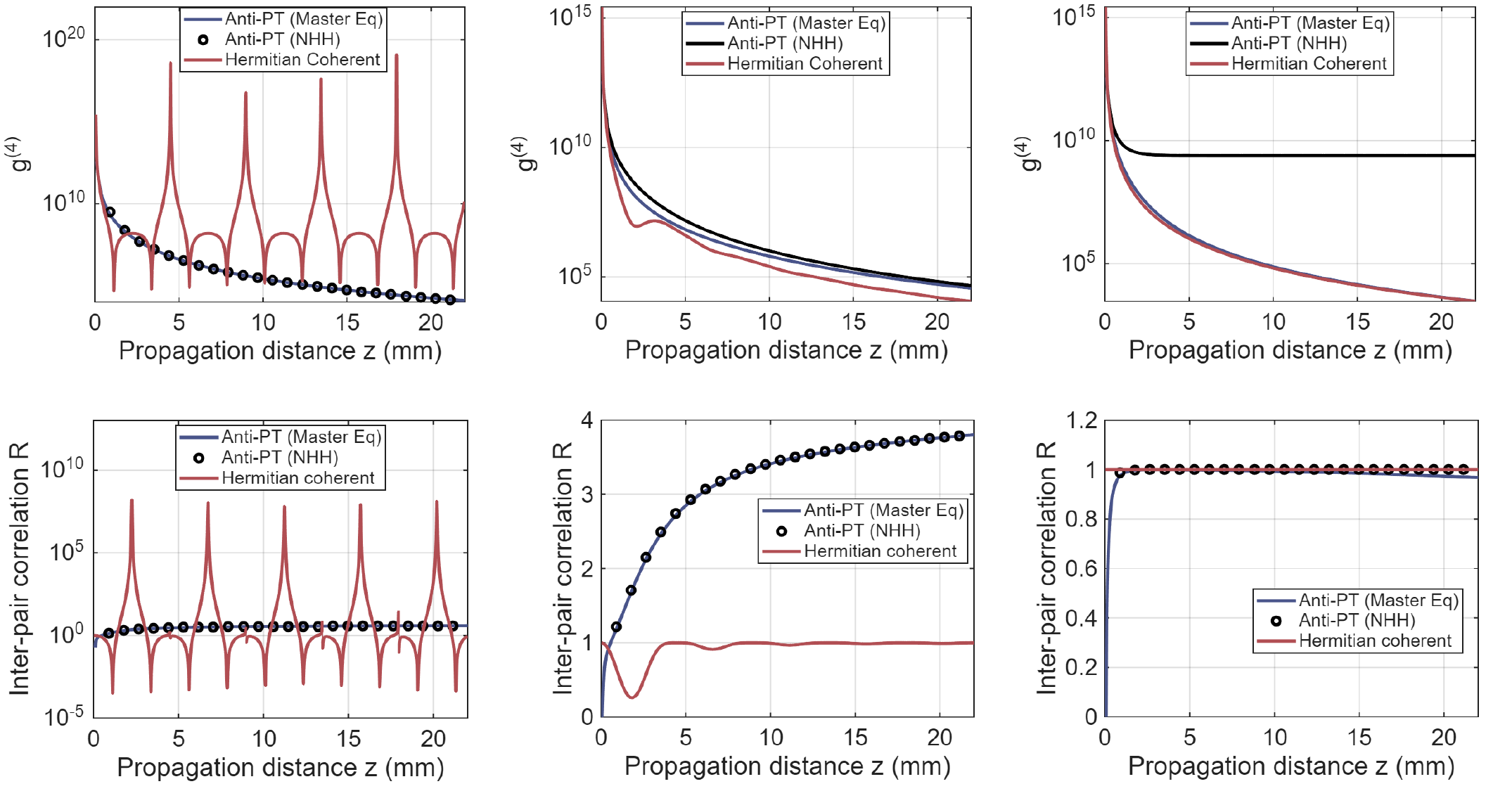}};

  \path[use as bounding box] (img.south west) rectangle (img.north east);
\node at (0.9,8.8) {\textbf{(a)}};
\node at (5.7,8.8) {\textbf{(b)}};
\node at (11.2,8.8) {\textbf{(c)}};
\node at (0.9,4.4) {\textbf{(d)}};
\node at (5.7,4.4) {\textbf{(e)}};
\node at (11.2,4.4) {\textbf{(f)}};
\end{tikzpicture}

\caption{\textbf{The effect of quantum jumps on four-photon correlations}. (a-c) Phase dependence of the normalized four-photon correlations $g^{(4)}$. For $\theta>0$, the $g^{(4)}$ of master equation and NHH approaches show noticeable discrepancies due to differences in the mean photon numbers. (d-f) The four-photon inter-pair correlation $\mathcal{R}^{(4)}$ exhibits only minor differences between the two approaches. The reference Hermitian coherent system is included for comparison.}\label{fig5}
\end{figure}
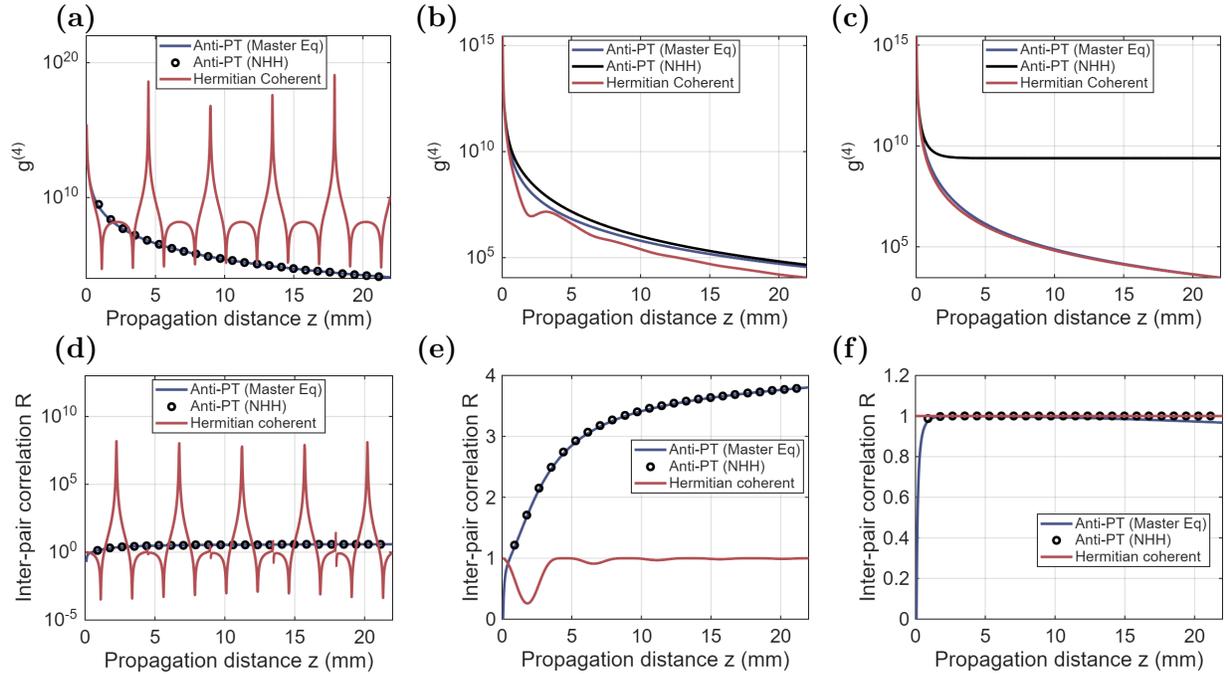

\section{Supplementary Section S3.~Effects of Quantum Jumps}
The quantum jumps $4\Gamma\sum_{\mu}B_{\mu}\rho B_{\mu}^{\dagger}$ lead to deviations in the mean photon number between the master-equation and non-Hermitian-Hamiltonian approaches for $\theta>0$, as shown in Fig. 3(b,c) in the main text. Consequently, quantities that are highly sensitive to the mean photon number, such as the normalized  $g^{(4)}$, exhibit noticeable discrepancies, as shown in Fig.~S4(a-c). In contrast, the ratio $\mathcal R^{(4)}$, which depends only on the unnormalized four- and two-photon correlations, is largely unaffected, as shown in Fig.~S4(d-f).
\begin{figure}[h]
\centering

\begin{tikzpicture}
\node[anchor=south west, inner sep=0] (img) at (0,0)
    {\includegraphics[width=0.8\linewidth]{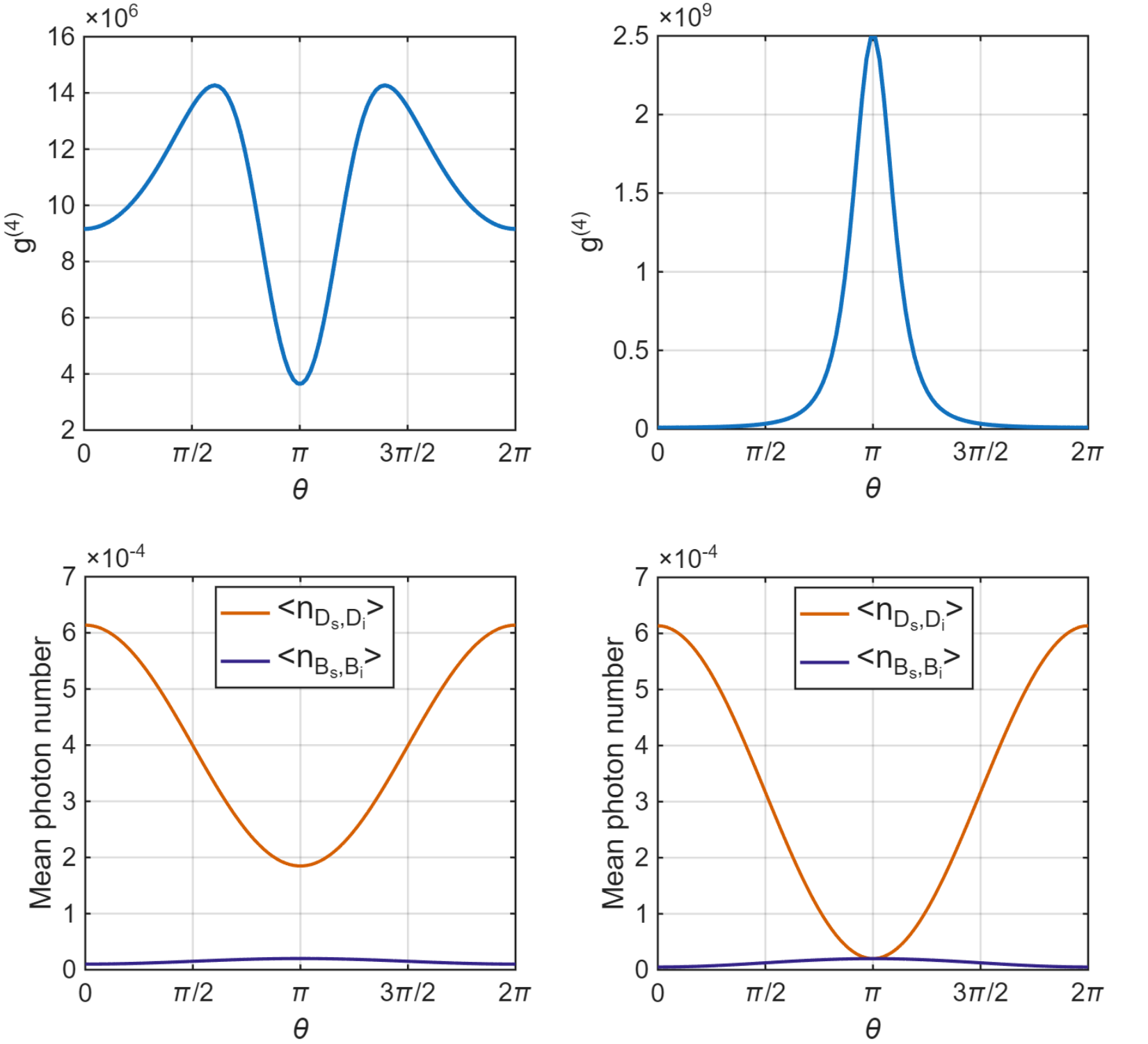}};

  \path[use as bounding box] (img.south west) rectangle (img.north east);
\node at (0.3,11.8) {\textbf{(a)}};
\node at (6.7,11.8) {\textbf{(b)}};
\node at (0.3,5.7) {\textbf{(c)}};
\node at (6.7,5.7) {\textbf{(d)}};
\end{tikzpicture}

\caption{\textbf{Phase dependence of quantum-jump effects on normalized correlations and mean photon numbers.} (a,b) The master-equation and non-Hermitian-Hamiltonian (NHH) approaches exhibit opposite phase dependences: at $\theta=\pi$, the master-equation result shows a minimum, whereas the NHH result shows a maximum. (c,d) Phase dependence of the mean photon numbers in the bright and dark modes for the two approaches.}\label{fig5}
\end{figure}
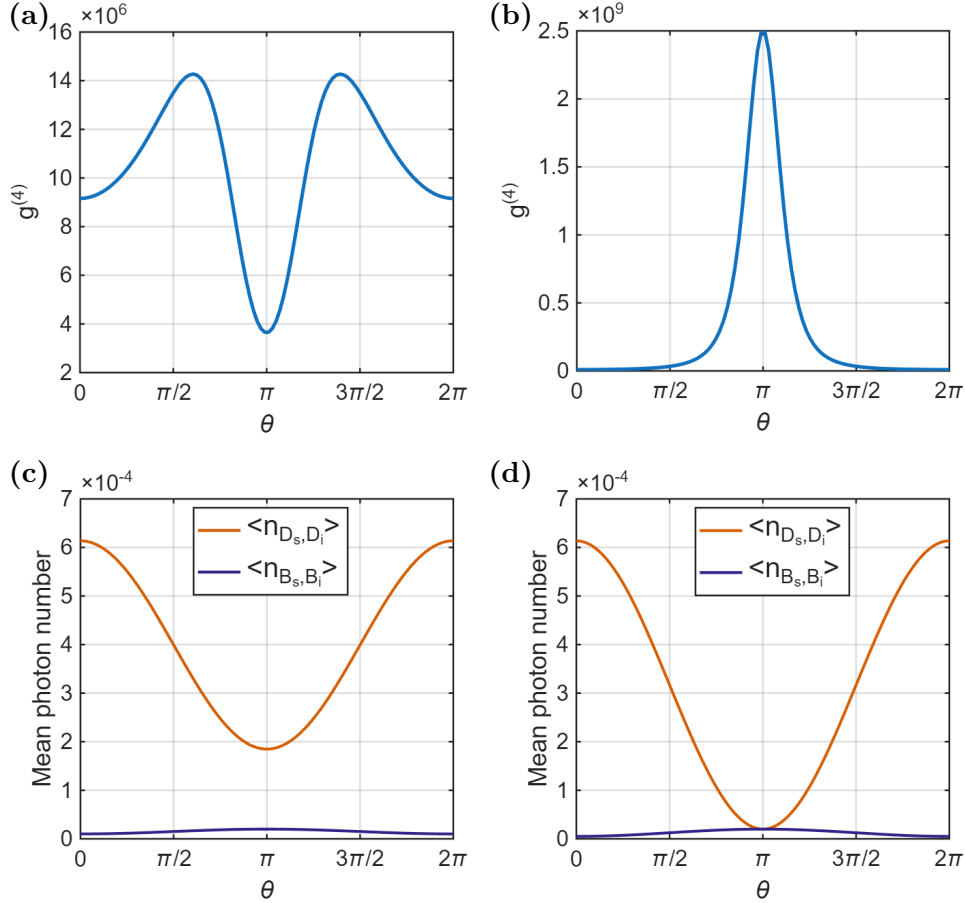
\begin{figure}[h]
\centering
\includegraphics[width=0.7\textwidth]{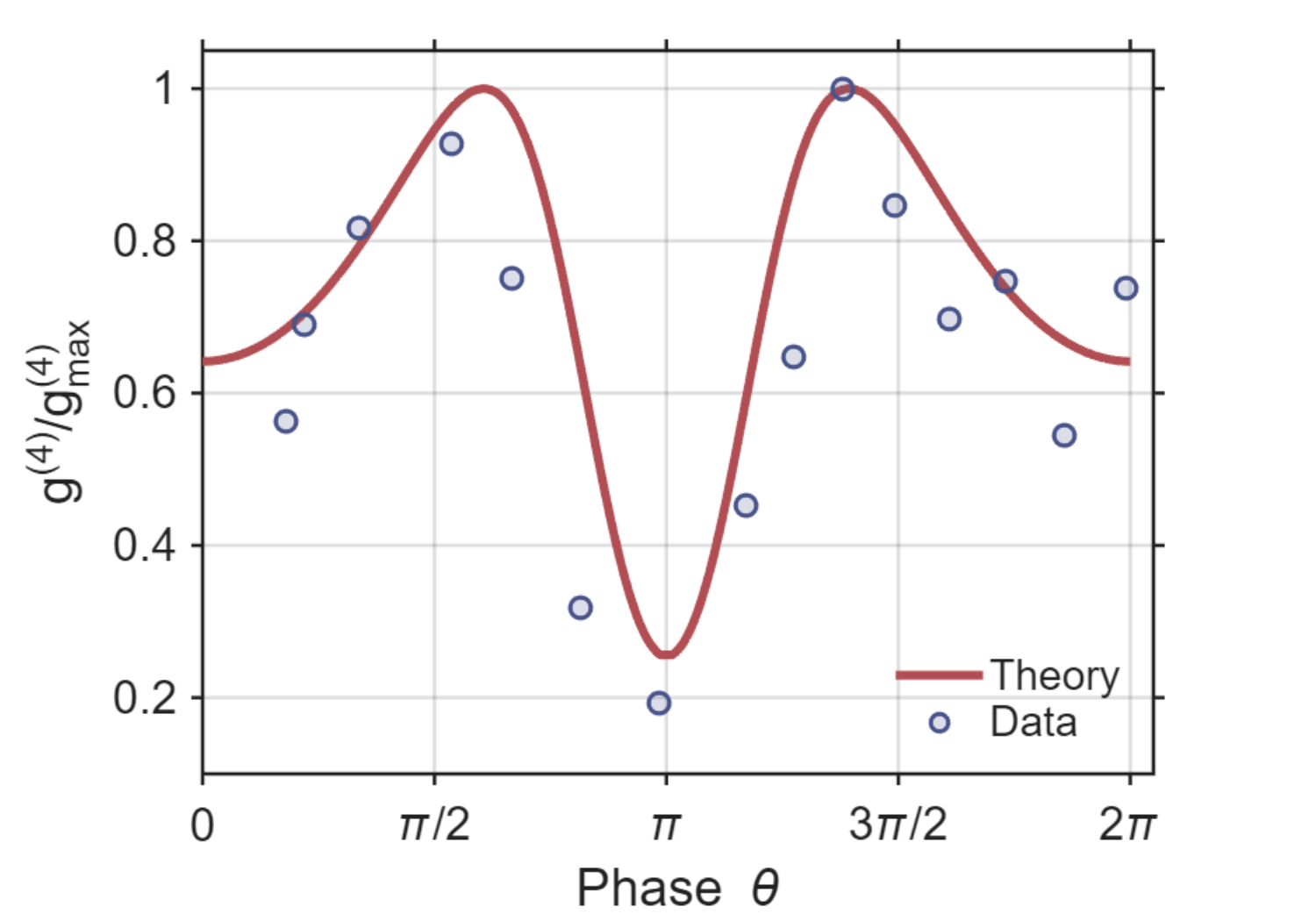}
\caption{\textbf{Experimental result for phase tuning of the normalized four-photon correlation} $g^{(4)}$. Both the experimental and theoretical results are normalized to their respective peak values.}\label{figS6}
\end{figure}
To understand this discrepancy in the mean photon numbers induced by quantum jumps, we plot the phase tuning of $g^{(4)}$ for both approaches, along with the corresponding phase dependence of the mean photon number in the bright/dark basis. As shown in Fig.~S5(b), the correlation exhibits a maximum at $\theta=\pi$, which is opposite to the behavior observed in the master equation approach as in Fig.~S5(a). We therefore focus on the specific case $\theta=\pi$, where only cross-correlated pairs are generated. We consider the action of the jump operator on an initial state $|n_{B_s},n_{B_i};n_{D_s},n_{D_i}\rangle$, which yields the modified density matrix
\begin{equation}
\begin{split}
        \rho_\mathrm{new}&=4\Gamma \sum_{\mu}B_{\mu}\rho B_{\mu}^{\dagger}=4\Gamma |n_{B_s}-1,n_{B_i};n_{D_s},n_{D_i}\rangle\langle n_{B_s}-1,n_{B_i};n_{D_s},n_{D_i}|\\
        &+4\Gamma|n_{B_s},n_{B_i}-1;n_{D_s},n_{D_i}\rangle\langle n_{B_s},n_{B_i}-1;n_{D_s},n_{D_i}|.
\end{split}
\end{equation}
When a system with $N$ photons loses a photon to the environment, it jumps to a state with $N-1$ photons. The resulting states $|n_B-1, n_D\rangle$ contain single dark photons originating from broken pairs—instances where the bright partner was dissipated. This accumulation of unpaired photons results in an unequal population of bright and dark states at $\theta=\pi$, as shown in Fig.~S5(c). In contrast, the effective Hamiltonian approach (without jumps) ignores these broken pair states, effectively post-selecting only the surviving correlated pairs. This results in an artificial equality in the population of bright and dark states at $\theta=\pi$, as shown in Fig.~S5(d). Because the single dark photons from broken pairs—which constitute noise—are excluded from the mean photon number in the effective Hamiltonian calculation, the denominator of $g^{(4)}$ is underestimated. Consequently, this approach predicts a higher, but unphysical, correlation value than that obtained from the master equation. The experimental measurement (normalized to maximum) of $g^{(4)}$ versus $\theta$ is shown in Fig.~S6. The $g^{(4)}$ reaches a minimum at $\theta=\pi$, confirming the theoretical prediction of the quantum jump effect.

\section{Supplementary Section S4.~Calibration of Phase Tuning}
\begin{figure}[h]
\centering
\includegraphics[width=0.8\textwidth]{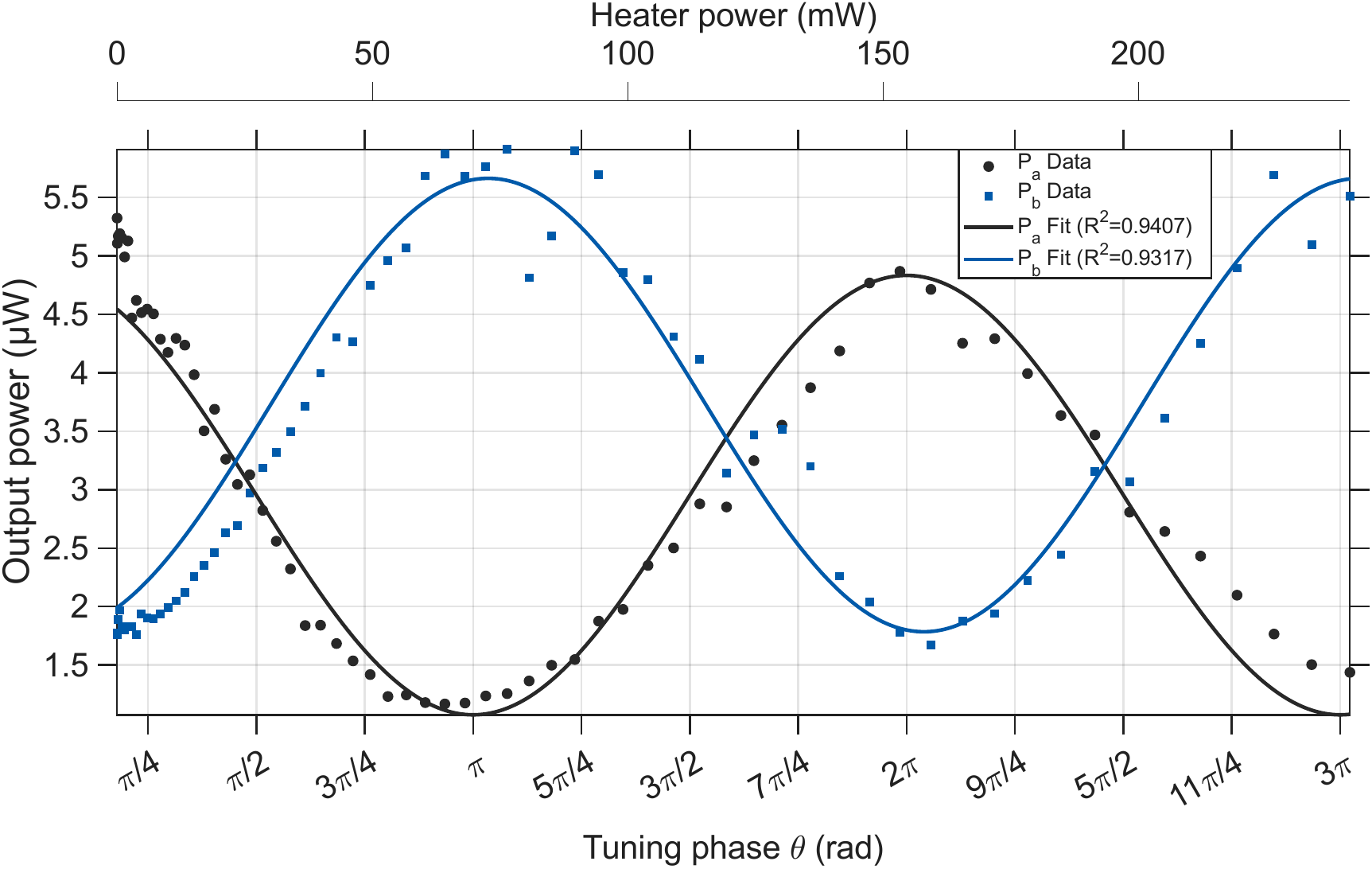}
\caption{\textbf{Calibration of tuning phase}. The classical output powers $P_a$ and $P_b$ from waveguides $a$ and $b$ are measured as the heater power is increased from 0 to 242 mW. Fitting the measured data yields a phase-conversion coefficient of $\beta=0.037$ rad/mW and an initial phase of $\theta_0=0.56$ rad. These calibrated parameters are used to map heater power to the corresponding phase in all phase-tuning measurements.}\label{fig3}
\end{figure}

 The tuning phase is calibrated by measuring the classical output power at the pump wavelength using CW pump injection. Since the condition for adiabatic elimination of the pump ($\gamma_{775} \gg \kappa_{775}$) is not met (see parameters in Section S1), the system functions approximately as three evanescently coupled waveguides for the classical pump light. Consequently, instead of exhibiting the equal output powers characteristic of the anti-PT system, the powers at the two waveguide outputs, $P_a$ and $P_b$, show complementary phase tuning curves \cite{Anti_PT2}, as displayed in Fig.~S7. The relationship between the relative phase $\theta$ and the heater power $P_\mathrm{heater}$ is mapped by fitting the data to 
 \begin{equation}
     P_{a,b} = \alpha \cos(\theta)+\eta,\ \ \ \theta = \beta P_{\text{heater}} + \theta_0,
 \end{equation}
where $\alpha$ and $\eta$ are fitting parameters representing the oscillation amplitude and the constant output power offset, respectively.
The fit yields a phase-conversion coefficient of $\beta=0.037$ rad/mW and an initial phase of $\theta_0=0.56$ rad.

\bibliography{citations}